\documentclass[12pt,reqno]{article}
\pdfoutput=1
\usepackage[latin1]{inputenc}
\usepackage{amsmath}
\usepackage{amsfonts}
\usepackage{amssymb}
\usepackage{graphicx}
\usepackage{geometry}
\usepackage{amssymb,epsfig}
\usepackage{hyperref}
\usepackage{subfig}
\usepackage{enumerate}

\makeatletter
\renewcommand\section{\@startsection {section}{1}{\z@}%
                                 {-3.5ex \@plus -1ex \@minus -.2ex}%nn
                                   {2.3ex \@plus.2ex}%
                                   {\normalfont\large\bfseries}}
\renewcommand\subsection{\@startsection{subsection}{2}{\z@}%
                                   {-3.25ex\@plus -1ex \@minus -.2ex}%
                                     {1.5ex \@plus .2ex}%
                                     {\normalfont\bfseries}}
\renewcommand\subsubsection{\@startsection{subsubsection}{3}{\z@}%
                                   {-3.25ex\@plus -1ex \@minus -.2ex}%
                                     {1.5ex \@plus .2ex}%
                                     {\normalfont\itshape}}
\makeatother

\def\pplogo{\vbox{\kern-\headheight\kern -29pt
\halign{##&##\hfil\cr&{\ppnumber}\cr\rule{0pt}{2.5ex}&\ppdate\cr}}}
\makeatletter
\def\ps@firstpage{\ps@empty \def\@oddhead{\hss\pplogo}%
  \let\@evenhead\@oddhead % in case an article starts on a left-hand page
}%      The only change in \maketitle is \thispagestyle{firstpage} instead of \thispagestyle{plain}
\def\maketitle{\par
 \begingroup
 \def\thefootnote{\fnsymbol{footnote}}
 \def\@makefnmark{\hbox{$^{\@thefnmark}$\hss}}
 \if@twocolumn
 \twocolumn[\@maketitle]
 \else \newpage
 \global\@topnum\z@ \@maketitle \fi\thispagestyle{firstpage}\@thanks
 \endgroup
 \setcounter{footnote}{0}
 \let\maketitle\relax
 \let\@maketitle\relax
 \gdef\@thanks{}\gdef\@author{}\gdef\@title{}\let\thanks\relax}
\makeatother

\numberwithin{equation}{section}

\newcommand{\be}{\begin{equation}}
\newcommand{\bea}{\begin{eqnarray}}
\newcommand{\ee}{\end{equation}}
\newcommand{\eea}{\end{eqnarray}}
\newcommand\beq{\begin{equation}}
\newcommand\eeq{\end{equation}}

\begin{document}
\setcounter{page}0
\def\ppnumber{\vbox{\baselineskip14pt
%\hbox{hep-th/0000000}
}}
\def\ppdate{}\date{}

\author{Sarah Harrison\\
 {\small \upshape\ttfamily sarharr@stanford.edu}\\
[7mm]{\normalsize \it Stanford Institute for Theoretical Physics }\\
{\normalsize  \it Department of Physics and SLAC, }\\
{\normalsize \it Stanford, CA 94309, USA}\\
[3mm]
}

\bigskip
\title{\bfseries Landau Levels, Anisotropy and Holography
\vskip 0.5cm}
\maketitle

\begin{abstract}
We analyze properties of field theories dual to extremal black branes in (4+1) dimensions with anisotropic near-horizon geometries. Such gravity solutions were recently shown to fall into nine classes which align with the Bianchi classification of real three-dimensional Lie algebras. As a warmup we compute constraints on critical exponents from energy conditions in the bulk and scalar two point functions for a general type I metric, which has translation invariance but broken rotations. We also comment on the divergent nature of tidal forces in general Bianchi-type metrics. Then we come to our main focus: extremal branes whose near-horizon isometries are those of the Heisenberg algebra (type II). We find hyperscaling-violating solutions with type II isometries in (4+1)-dimensions. We show that scale invariant (4+1)-dimensional type II metrics are related by Kaluza-Klein reduction to more symmetric $AdS_2\times R^2$ and (3+1)-dimensional hyperscaling-violating spacetimes. These solutions generically have $\theta \leq 0$. We discuss how one can obtain flows from UV CFTs to Bianchi-type spacetimes in the IR via the Higgs mechanism, as well as potential inhomogeneous instabilities of type II. Finally, we compute two-point functions of massive and massless scalar operators in the dual field theory and find that they exhibit the behavior of a theory with Landau levels.

\end{abstract}
\bigskip
\newpage

\tableofcontents

%%%%%%%%%%%%%%%%%%%%%%%%%%%%%%%%%%%%%%%%
%%%%%%%%%%%%%%%%%%%%%%%%%%%%%%%%%%%%%%%%
%%%%%%%%%%%%%%%%%%%%%%%%%%%%%%%%%%%%%%%%
%%%%%%%%%%%%%%%%%%%%%%%%%%%%%%%%%%%%%%%%
\section{Introduction}\label{sec:intro}

The AdS/CFT correspondence is a useful tool for studying aspects of strongly coupled quantum field theories \cite{magoo}. The first and most precisely developed form of gauge/gravity duality relates conformal field theories in $d+1$ dimensions to classical gravity in $d+2$-dimensional AdS space. However, due to novel results of applications of holography to condensed matter physics \cite{sean, chris, john,subir}, recently there has been a flurry of interest in applying gauge/gravity duality to  systems which are not conformally invariant. After all, most real world materials exhibit a plethora of phases which do not obey conformal invariance. Motivated by these observations, holography has been extended to Lorentz-symmetry breaking field theories which obey dynamical scaling \cite{KLM,Taylor,GKPT} and to non-scale-invariant theories which exhibit hyperscaling violation\cite{thetaglom,edgar,theta}. However, these theories are still highly symmetric, enjoying translational and spatial rotational isometries.

Extremal black branes in the bulk correspond to zero temperature ground states in the doped field theory. Until very recently, we have only had few examples of such extremal brane solutions, the simplest of which is the AdS-Reissner-Nordstrom black brane. This solution is known to have a near horizon $AdS_2\times R^d$ region with an extensive ground state entropy, and for this reason it is widely speculated that it is unstable and that some other geometry represents the true end to the RG flow. In fact, given our knowledge of the vast number of zero-temperature states in condensed matter theory, the true field theory ground state could likely be both anisotropic and inhomogeneous. From the gravity point of view, these field theories are dual to geometries with anisotropic, inhomogeneous extremal black branes embedded in asymptotically $AdS$ space. Besides the fact that finding such extremal black brane solutions in gravity is interesting in its own right, given our knowledge of the vast ``landscape" of exotic (including inhomogeneous and anisotropic) phases of matter, the prospect of finding and classifying these solutions is tantalizing from the dual field theory perspective. 

The results of \cite{bigteam} are a first step towards doing just this: Iizuka et al. provide a classification of homogeneous, anisotropic black brane solutions in 4+1 dimensions which is based on the Bianchi classification of homogeneous, anisotropic cosmologies. These near-horizon solutions can be viewed as attractors from the bulk perspective, which means that although the microscopics of a particular theory may dictate which solutions are fixed points of the RG flow, the properties of the fixed points are universal, and we can analyze them only using the near-horizon attractor solution. In fact, the authors of \cite{bigteam2} extend their solutions to hyperscaling-violating and inhomogeneous phases for a particular subset of their solutions. For other discussions of inhomogeneous phases in holography, see \cite{ooguri,modulation,santos,vlattice}. 

In this paper, we take a closer look at extremal black branes with the isometries of the type II algebra--the Lie algebra of the three-dimensional Heisenberg group. This case is particularly interesting because of its simple interpretation as a magnetic field in  four dimensions which comes from KK reduction of one spatial direction. Thus it may be relevant if one would like to find gravity duals of charged fluids in a magnetic field, which exhibit a variety of fascinating phenomena, such as the fractional quantum Hall effect. 

The outline of this paper is as follows. In the rest of the introduction, we briefly review the classification of homogeneous anisotropic black brane geometries discussed in \cite{bigteam}, which makes use of Bianchi's classification of three-dimensional real Lie algebras. In \S~\ref{sec:type1} we discuss properties of type I spacetimes which have three-dimensional translations but with broken rotational symmetry. We discuss constraints on critical exponents from energy conditions in the bulk (including the possibility of hyperscaling violation,) and we compute two point functions for massive scalars in the semiclassical approximation. Finally, we end this section with a brief interlude on tidal forces in the homogeneous, anisotropic spacetimes of \cite{bigteam}. As one would expect, due to spatial anisotropy, the tidal forces are also anisotropic (and divergent), however, for some particular values of critical exponents, they may be rendered finite.

We move on to the main focus of this paper in \S~\ref{sec:heis}, spacetimes with isometries of the three-dimensional Heisenberg group. We discuss the forms of the Heisenberg algebra and its representations, and we compute geodesics in (3+1)-dimensional spacetimes with this symmetry group (before the addition of a ``radial direction.") We find that massive particles (classically) follow spiral trajectories, just as an electrically charged particle in a constant magnetic field. 

We move onto holography in \S~\ref{sec:scaling}, where we add a radial direction and once again compute constraints on critical exponents due to energy conditions in the bulk. We show that the scale invariant (4+1)-dimensional type II solution can be generalized to one in which the metric has an overall hyperscaling violation exponent; this only requires adding a log-running dilaton field to the action, and can be achieved for a reasonable range of parameters. We also show that one can achieve $AdS_2\times R^2$ and (3+1)-dimensional hyperscaling-violation (generically with $\theta\leq 0$) with fully-restored translation invariance in the bulk through KK reduction along the direction of the ``magnetic field." The former are known to exhibit inhomogeneous phases \cite{vlattice}, whereas the latter are known to arise in the near-horizon region of D2 branes with a compact $S^6$. We also compute massive scalar correlators in the semiclassical limit and briefly discuss entanglement in five-dimensional type-II spacetimes and their KK-reduced four-dimensional theories.

Finally, in \S~\ref{sec:2pt}, we compute two-point functions for operators dual to bulk scalar fields in the scale-invariant type II background by solving the bulk equations of motion (for both massless and massive scalars in the probe limit.) Call $y$ the direction of the KK magnetic field and $(x,z)$ the coordinates in the plane threaded by the magnetic field. We find the scalar two-point function has the form of a harmonic oscillator coherent state, 
\be
\langle\mathcal O(\tau,x,z,k_y)\mathcal O(\tau',x',z',-k_y)\rangle\sim\frac{1}{|\Delta\tau|^{2\Delta(k_y)}}f(k_y)e^{-{k_y\over 4}(\Delta x^2+\Delta z^2-2i(x+x')|\Delta z|)}
\ee
as is characteristic of the wave function of an electron in Landau levels with $y$-momentum-dependent charge and scaling dimension. We find, as should be expected, that (up to a phase) it is independent of the gauge we choose for the bulk metric. In four-dimensions, $k_y$ can only take on discrete values, and this two-point function thus describes a charged scalar coupled to the four-dimensional magnetic field, with charge given by $k_y$, its momentum along the compact direction. The advantage of using the five-dimensional metric for this computation is that we can see the structure of Landau levels in the dual field theory through a computation that only relies on the metric and a neutral scalar in the probe approximation. We also comment on the choice of state in the dual field theory. In \S~\ref{sec:discussion} we conclude with a discussion of our results and possible future directions.
 
\subsection{Review of Bianchi attractor classification}\label{sec:rev}
All three dimensional real Lie algebras were classified by Bianchi. Each class has three infinitesimal generators that form a closed algebra, and is defined by the algebra's structure constants. These generators obey commutation relations of the form
\be
[\xi_i,\xi_j]=C^k_{ij}\xi_k,
\ee
where the structure constants, $C^k_{ij}$, are real. There are nine classes of algebras of this form defined by nine possible sets of structure constants (up to trivial redefinition, such as change of basis.) Each generator of the Lie algebra is a basis vector in a vector space of infinitesimal coordinate transformations. These generators will be Killing vectors of any metric which obeys these symmetries. Such spaces are homogeneous, as each point has the same infinitesimal tangent space. However, in general the Killing vectors do not commute; they break three-dimensional translation invariance, and the space is anisotropic. 

First, how can we construct metrics with these spatial symmetries? Two observations are useful. The first is that there is a set of three vector fields, $X_i$, which commute with each generator of the algebra and which obey the commutation relations,
\be
[X_i,\xi_j]=0,~~[X_i,X_j]=-C^k_{ij}X_k.
\ee
The second observation is that  tangent space (generated by the $\xi_i$) at each point is dual to a cotangent space generated by a set of three one-forms. There are three one-forms, $\omega_i$, in the cotangent space which are invariant under the Killing isometries generated by the $\{\xi_i\}$; these are the one forms dual to the vector fields $X_i$ which commute with the Killing vectors. They obey the relation,
\be
d\omega^i=C^i_{jk}\omega^j\wedge \omega^k.
\ee
A metric constructed from wedge products of these one forms will also be invariant under the isometries generated by the Killing vectors.

We can then add scaling by adding time (this adds an additional killing vector $\partial_t$) and an extra spatial dimension, the ``radial" coordinate, leading to the general form \cite{bigteam}
\be\label{eq:gen_metric}
ds^2=L^2\left [{dr^2\over r^2}-r^{-2\beta_t}dt^2+\eta_{ij}r^{-(\beta_i+\beta_j)}\omega^i\otimes \omega^j\right ]
\ee
where $r\to 0$ is the boundary, $\eta_{ij}$ is a constant matrix (for simplicity taken to be diagonal), and the $\omega^i$ are the one forms which are invariant under the isometries of the Lie algebra. In the metric of equation (\ref{eq:gen_metric}), there is an additional Killing vector which implements a scaling symmetry via translation along the radial direction $r\to r+\epsilon$.  We have also allowed for general scaling exponents in front of each of the invariant one-forms. Note that our conventions are different from those in \cite{bigteam,bigteam2}; their boundary was defined as $r\to\infty$, whereas ours is defined as $r\to 0$.

In \cite{bigteam} it was shown that metrics for most of the three-dimensional Bianchi classes can be supported by an action of the form, 
\be\label{eq:gen_action}
S=\int d^5\sqrt{-g}\left (R-2\Lambda-{1\over 4}F^2-{1\over 4}m_i^2A_i^2\right ),
\ee 
with one or two massive vector fields, $A_i$.
These metrics can be interpreted as the near-horizon region of an extremal black brane geometry, and are effective only up to some radial scale $r_F$. We expect that the solution may cross over to one with conformal invariance in the UV\footnote{In fact, for many of the solutions found, $A\sim r^{\gamma}$ for $\gamma <0$, which grows in the UV. Clearly this divergent solution cannot continue all the way to the boundary, $r=0$.}, as $r\to 0$, and that for $r\gg r_F$ there may be some other type of fixed point in the IR. There may also be other intermediate fixed points, such as translationally-invariant hyperscaling violating or Lifshitz solutions at other radial scales. As in \cite{theta}, we take an effective approach, assuming the metric of equation (\ref{eq:gen_metric}) is valid at some intermediate radial region, and we evaluate quantities at a fixed radial slice.
 
\section{Warmup: General Scaling in Bianchi Type I}\label{sec:type1}
Before we come to the physics of spacetimes with type II isometries, we discuss some salient aspects of the most general type I metric, of which both Lifshitz and AdS are rotationally invariant special cases. In \S~\ref{sec:t1NEC} we compute constraints on the values of critical exponents coming from the Null Energy Conditions in the bulk. In \S~\ref{sec:typeIscaling} we comment on the expected scaling of a general scalar two-point function in type I, while in \S~\ref{sec:typeIgeodesic} we compute two-point functions of massive scalar operators in the dual field theory in the semiclassical limit.  Finally, in \S~\ref{sec:tidal} we compute the tidal forces on a massive particle falling radially in a general Bianchi-type metric, and show that they are generically anisotropic and divergent, though for particular combinations of scaling exponents, they may be rendered finite.
\subsection{Curvature and energy conditions}\label{sec:t1NEC}
The isometries of the type I algebra are the usual commuting translations of $\mathbb R^3$, with Killing vectors $\partial_x,\partial_y$ and $\partial_z$. However, in five bulk dimensions this algebra allows for a more general class of metrics than pure (conformally) AdS and Lifshitz spacetimes due to the fact that the three spatial directions can scale anisotropically with the radial coordinate. In this case, three-dimensional rotational invariance in R$^3$ is broken. The most general scale-invariant (or scale-covariant, if $\theta\neq 0$) metric takes the form
\be\label{eq:typeImet}
ds^2=L^2r^{2\theta\over 3}(-r^{-2\beta_t}dt^2+{dr^2\over r^2} +r^{-2\beta_x}dx^2+r^{-2\beta_y}dy^2 +{dz^2\over r^2}),
\ee
where the isometries allow for general values of the $\beta_i$. Upon rescaling $r\to\lambda r$, the coordinates will transform as $x_i\to\lambda^{\beta_i} x_i$ for $x_i\in (t,x,y,z)$, and the metric scales with an overall conformal factor, $ds\to \lambda^{\theta/3}ds$. (We choose this convention to match that of \cite{theta}.) Note that we have set the $z$ coordinate to scale with weight 1, which we can always do through a redefinition of $r$ (as long as $\beta_z\neq 0$.) In the case of $\theta=0$, the scalar curvature is constant and the metric is scale-invariant.

We would like to know what values of parameters in equation (\ref{eq:typeImet}) yield physically realizable metrics. One such test is that the metric satisfies reasonable energy conditions, such as the Null Energy Condition, $T_{\mu\nu}N^\mu N^\nu\geq 0$, where $N^\mu$ is a null vector with respect to the metric of equation (\ref{eq:typeImet}). This will constrain the values of the free parameters in the metric, and thus the values of the critical exponents in the dual field theory. Assuming the metric satisfies the Einstein equations, $G_{\mu\nu}=T_{\mu\nu}$, we can compute $G_{\mu\nu}N^\mu N^\nu$ to get constraints on $\theta$ and $\{\beta_i\}$.

It turns out that the allowed combinations of critical exponents are highly constrained. The full NEC are given in Appendex \ref{sec:NEC}, but here we mention a few particular limits. 
\begin{itemize}
\item $\beta_x=\beta_y=1$

This case reduces to five-dimensional Lifshitz with hyperscaling violation. The NEC for this can be found in, e.g., \cite{theta}. In fact, we would get similar conditions on $\theta, \beta_i$ to that of a Lifshitz solution if $\beta_t=\beta_{j\neq i}=1$ for all but one spatial direction.
\item $\beta_t=1, \theta=0$

In this case, we get conditions on the spatial scaling exponents. If $\beta_x=0$ or $\beta_x=1$ then $0\leq \beta_y\leq 1$. The NEC are symmetric in $\beta_x,\beta_y$, so the same applies if we switch them. If we set $\beta_x=\beta_y=\beta$, then the limits on $\beta$ are $0\leq \beta\leq 1$. Note that even though $\beta_t=1$, we are still breaking Lorentz invariance if one of the $\beta_i\neq 1$, as the spacetime anisotropy breaks both rotations and boosts.
\item $\beta_y=1,\theta=0$

In this case we allow for nontrivial scaling in the time and one of the spatial directions. The constraint is $\beta_t\geq \max(1,\beta_x)$ as long as $\beta_x\geq 0$.
\item $\beta_t=1,\beta_y=0,\theta\neq 0$

In this case we have the conditions $\beta_x\leq 1$, $\beta_x\geq \theta-2$, and $\beta_x(1-\beta_x)-\theta+\theta^2/3\geq 0$. If $\beta_x=1$, then $\theta=3$ (this essentially corresponds to flat space) or $\theta\leq 0$.
\item $\beta_t=1,\beta_x=\beta_y=\beta, \theta\neq 0$

Here we have the conditions $2\beta(1-\beta)\geq \theta-\theta^2/3$ and $(1-\beta)(2+2\beta-\theta)\geq 0$. These reduce to $\theta\geq 3$ or $\theta \leq 0$ if $\beta=1$.
\end{itemize}
Note that the metric of equation (\ref{eq:typeImet}) will in general only be valid for a range of radial scales, as discussed in \S~\ref{sec:rev}. It is possible that some of the conditions above may have instabilities though they satisfy the NEC. For instance, the conditions which allow $\theta >3$ are likely unstable generically.\footnote{It was argued in \cite{theta} that when the hyperscaling violation exponent takes on a value greater than $d$, where $d$ is the spatial dimension of the dual field theory, the theory is unstable.} 

\subsection{Scaling of two-point functions with spatial anisotropy}\label{sec:typeIscaling}
Before we do any computation, let's discuss the expected form of two-point functions of scalar operators (in the energy regime where equation (\ref{eq:typeImet}) is valid) based on scaling arguments. We will consider a bulk scalar field with action
\be
S=\int d^5x\sqrt{-g}( -(\partial_\mu \phi)^2-m^2\phi^2),
\ee
which couples to a scalar operator $\mathcal O(t,\vec x)$ in the dual field theory through a term
\be\label{eq:coupling}
\int d^4x\phi_0(t,\vec x)\mathcal O(t,\vec x)
\ee
on the boundary, where $\phi_0(t,\vec x)$ is the boundary value of the bulk scalar field. From the usual prescription of AdS/CFT, the $\langle \mathcal O\mathcal O\rangle$ two-point function in the field theory is then given by the second derivative of the action with respect to the boundary value of the bulk field,
\be\label{eq:2ptfunc}
\langle \mathcal O(t,\vec x)\mathcal O(t',\vec x')\rangle=\frac{\delta^2}{\delta \phi_0(t,\vec x)\delta \phi_0(t',\vec x')}S\bigg|_{r\to \epsilon}.
\ee
The field $\phi$ will have an expansion near the boundary,
\be
\phi(r,\vec k)\sim 1+\ldots + r^{1+\beta_t+\beta_x+\beta_y-\theta}G(\vec k,m)
\ee
where $G(\vec k,m)$ is the momentum-space two-point function.
In order to preserve scale-invariance, this implies that if we rescale $r\to\lambda r$,
\be
G(\omega, k_x,k_y,k_z,m)\to \lambda^{1+\beta_t+\beta_x+\beta_y-\theta}G(\omega/\lambda^{\beta_t}, k_x/\lambda^{\beta_x},k_y/\lambda^{\beta_y},k_z/\lambda,m/\lambda^{\theta/3}).
\ee
In general this has complicated momentum dependence; we will see more detailed examples of this when we consider two-point functions of scalars in a type II background in \S~\ref{sec:LL2pt}. As a check, we see that our results reduce to the rotationally-invariant case\cite{theta} for $\beta_x=\beta_y=1$.
\subsection{Geodesic approximation}\label{sec:typeIgeodesic}
Here we will compute two-point functions of an operator coupled to a probe scalar field in a general type I background in the semiclassical limit; this will also become useful when we return to spacetimes with Heisenberg isometries. In this section we will stick to the case of $\theta=0$.

The correlation function $\langle \mathcal O(t_i,\vec x_i)\mathcal O(t_f,\vec x_f)\rangle$, where $\mathcal O$ is some massive operator of the boundary theory which couples to $\phi(r,t,\vec x)$ in the bulk through the term (\ref{eq:coupling}), can be computed in the semiclassical approximation by considering the geodesic of a massive particle which travels between two points on the boundary, $(\epsilon,t_i,\vec x_i)$ and $(\epsilon, t_f,\vec x_f)$, where $\epsilon$ is a UV regulator. The geodesic is computed by extremizing the action for a particle which travels between these two points, and the propagator is given by 
\be
\langle \mathcal O(t_i,\vec x_i)\mathcal O(t_f,\vec x_f)\rangle\sim e^{S(x_i,x_f)}
\ee
where $S(x_i,x_f)$ is the action evaluated along the geodesic between $x_i$ and $x_f$.

First let's remind ourselves of two limiting cases of type I, $AdS_2\times R^3$ and $AdS_5$, each of which has both three-dimensional translation and rotation invariance. In the case of $AdS_2\times R^3$, the $r$ direction and the spatial directions decouple---therefore a geodesic will not travel into the bulk, and the action is simply given by the geodesic distance on the boundary, which is flat space; i.e. $S=-m\sqrt{\Delta x^2+\Delta y^2+\Delta z^2}$.
The propagator has the familiar exponential decay with distance,
\be\label{eq:r3G}
G\sim e^{-m\sqrt{\Delta x^2+\Delta y^2+\Delta z^2}},
\ee
which, because of translation invariance, only depends on the separation between the two points, $\Delta \vec x$. On the other hand, in the case of $AdS_5$, a geodesic will travel into the bulk in order to minimize its length; in this case the extremal action becomes $S\sim -2m\log\left (\frac{|\Delta \vec x|^2}{\epsilon^2}\right )$, which yields a boundary correlation function with the familiar power law behavior of conformally invariant theories,
\be
G\sim \left (\frac{\epsilon}{|\Delta \vec x|}\right )^{2m}.
\ee

What about generic type I spacetimes with anisotropic scaling in both the time and spatial directions? The action for a massive scalar field in the background metric of equation (\ref{eq:typeImet}) is
\be\label{eq:typeIaction}
S=-m\int d\lambda\sqrt{r^{-2\beta_t}\dot\tau^2+\frac{\dot r^2}{r^2}+r^{-2\beta_x}\dot x^2+r^{-2\beta_y}\dot y^2 +r^{-2}\dot z^2}
\ee
where $\lambda$ is an affine parameter along the geodesic, and we have Euclideanized by setting $t=i\tau$. We need to compute the value of this action at its saddle point. Making the choice $\lambda=r$, there are four conserved momenta,
\be
\pi_i=\frac{r^{-2\beta_i}x_i'}{\sqrt{r^{-2\beta_t}\tau'^2+r^{-2}+r^{-2\beta_x} x'^2+r^{-2\beta_y} y'^2 +r^{-2} z'^2}},
\ee
where $x_i'=\partial x_i/\partial r$. We can recast these equations in terms of the momenta as
\be
x_i'^2=\frac{\pi_i^2r^{4\beta_i}}{r^2\left(1-\sum\limits_j\pi_j^2r^{2\beta_j}\right )},
\ee 
where the sum runs over $j\in \{t,x,y,z\}$.
The geodesic will have an extremum in the bulk at a critical value of the radius, $r=r_c$, defined by $\partial x_i/\partial r |_{r=r_c}=0$ for all $x_i$. This imposes an additional constraint, $\sum\limits_j\pi_j^2r_c^{2\beta_j}=1$. Finally, we can get relationships between $\{|\Delta x_i|\}$, $r_c$, and $\{\pi_i\}$ by integrating the differential equations along the geodesic:
\be\label{eq:d1}
\frac{|\Delta x_i|}{2}=\int_\epsilon^{r_c} dr \frac{\pi_ir^{2\beta_i-1}}{\sqrt{1-\sum\limits_j\pi_j^2r^{2\beta_j}}},
\ee
and the action:
\be\label{eq:d2}
-\frac{S}{2m}=\int_\epsilon^{r_c} dr \frac{1}{r\sqrt{1-\sum\limits_j\pi_j^2r^{2\beta_j}}}.
\ee
This is not solvable analytically for generic values of the parameters; however, we will mention a few particularly interesting cases for which we can solve the equations. If the boundary points are separated in only one direction which scales with exponent $\beta_i$ (or equivalently, in multiple directions which all scale with the same exponent,) then solving for the separation $|\Delta x_i|$ between these points and plugging back into the action yields
\be
S=-2m\int_\epsilon^{r_c} dr \frac{1}{r\sqrt{1-\left({r\over r_c}\right)^{2\beta_i}}}=\frac{2m}{\beta_i}\tanh^{-1}\left (\frac{\sqrt{r_c^{2\beta_i}-\epsilon^{2\beta_i}}}{r_c^{\beta_3}}\right )=\frac{m}{\beta_i}\log\left (\frac{\beta_i^2|\Delta x_i|^2}{\epsilon^{2\beta_i}}\right )
\ee
The massive propagator then goes like
\be\label{eq:powerlawG}
G(|\Delta x_i|)\sim e^{S}\sim\left (\frac{ \epsilon}{|\Delta x_i|}\right )^{2m/\beta_i}.
\ee
This is the familiar power law form of a two-point function for points separated along a direction that scales with weight $\beta_i$. Setting $x_i=\tau$, we recover the dynamical scaling form of Lifshitz. It is clear that when $\beta_i=1$ this reduces to the result for a scalar in $AdS_5$.

Another case for which equations (\ref{eq:d1}) and (\ref{eq:d2}) can be solved is when the boundary points are only separated in two directions, say $x$ and $y$, and one of them doesn't scale, e.g. $\beta_x=0$. Then the action becomes
\be
S=-m\sqrt{|\Delta x|^2+\frac{\log^2\left ({\beta_y^2|\Delta y|^2\over \epsilon^{2\beta_y}}\right )}{\beta_y^2}},
\ee
which yields a propagator of the form
\be\label{eq:mixedG}
G(|\Delta x|,|\Delta y|)\sim e^{-m\sqrt{|\Delta x|^2+\frac{\log^2\left ({\beta_y^2|\Delta y|^2\over \epsilon^{2\beta_y}}\right )}{\beta_y^2}}}.
\ee
This has mixed exponential and power law behavior. Note that in the limit $\Delta y\to \epsilon$ we get a pure exponential and when $\Delta x\to 0$, the propagator becomes
\be
G\sim\left ( {\epsilon\over |\Delta y|}\right )^{-2m/\beta_y}
\ee
just as in equation (\ref{eq:powerlawG}), as we would expect. 

This is only valid at large mass and/or separation measured in terms of the radial cutoff $r_F$, i.e. in the IR regime where the mass term is important and $m|\Delta x_i|\gg 1$. 

\subsection{Tidal Forces in general Bianchi-type metrics}\label{sec:tidal}
Particles traveling radially in spacetimes with anisotropic scaling (such as Lifshitz type metrics) are known to experience divergent tidal forces as they approach the horizon despite all curvature invariants being finite\cite{edgar,tidal}. We expect this to be the case as well for generic Bianchi-type metrics. Even though all scale-invariant metrics of the form (\ref{eq:gen_metric}) have a constant  Ricci scalar, particles traveling along radial geodesics may experience tidal forces which diverge. When we allow for general anisotropic scaling, we also expect that these tidal forces will be spatially anisotropic.

The quantity which captures the behavior of tidal forces on particles traveling along geodesics is the geodesic deviation
\be
T^\nu\nabla_\nu (T^\mu\nabla_\mu\hat N),
\ee
where $T_\mu$ is the tangent vector along the geodesic and $\hat N$ is a unit normal vector to the geodesic. This quantity measures the relative acceleration of neighboring geodesics along the direction of the normal vector, $\hat N$. In general, this is spatially-dependent.

We will consider a massive particle propagating radially in a metric of the form of equation (\ref{eq:gen_metric}). This particle's trajectory can be found by extremizing the action
\be
S=-mL\int d\tau\sqrt{{\dot t^2\over r^{2\beta_t}}-{\dot r^2\over r^2}},
\ee
where $\tau$ is an affine parameter along the geodesic. There is a conserved energy $E$ conjugate to $\dot t$, $E=\frac{mL}{r^{2\beta_t}}\dot t$, and the action is solved by the trajectory
\be
\dot t=\frac{E}{mL}r^{2\beta_t};~~~\dot r=\frac{r}{mL}\sqrt{E^2r^{2\beta_t}-m^2}.
\ee
We would like to compute the geodesic deviation along this trajectory for particles spatially separated by $\Delta \vec x$. The tangent vector to this trajectory will  be given by $T^\mu=(\dot t,\dot r,0,0,0),$ and we will consider normal vectors of the form $N^\mu=(0,0,\hat x)$ where $\hat x$ is a unit vector along the $x_i$-direction, $i=1,2,3.$  The case of each normal vector must be considered separately because of the anisotropy in the metric.

For generic Bianchi-type metrics, the geodesic deviation along the $x_i$-direction takes the form
\bea
T^\nu\nabla_\nu (T^\mu\nabla_\mu\hat N)&\sim& \frac{E^2}{m^2L^2}\beta_i(\beta_i-\beta_t)r^{2\beta_t},~~\beta_i\neq 0\\
&\sim& 0,~~~~~~~~~~~~~~~~~~~~~~~~~~~\beta_i=0,
\eea
where $\beta_i$ is the exponent of $r$ in front of $dx_i^2$. This diverges for $\beta_i\neq \{0, \beta_t\}$ as the particle approaches the horizon, $r\to\infty$. Notice that, unlike in pure Lifshitz metrics, the appearance of divergent tidal forces is no longer just a function of the scaling exponent $\beta_t$.  In fact, we can break Lorentz symmetry and still have tidal forces which do not diverge in some directions; this will be the case in the directions $x_i$ which scale with weight $\beta_i=\beta_t$. Besides this, we should also mention that there can be additional coordinate-dependent contributions to the tidal forces if the metric is spacetime-dependent.

It is quite simple to generalize this to Bianchi-type metrics with nonzero $\theta$. In this case the geodesic deviation along the $x_i$-direction takes the form
\be
T^\nu\nabla_\nu (T^\mu\nabla_\mu\hat N)\sim\frac{3\beta_i-\theta}{L^2}\left (r^{2\beta_t-4\theta/3}(3\beta_t-3\beta_i-\theta)+3\beta_ir^{-2\theta/3}\right).
\ee
The first term in this equation diverges as $r\to\infty$ for generic values of the critical exponents. However, the tidal forces will be finite in any direction $x_i$ if $\beta_i=\theta/3$ or $\beta_i=\beta_t-\theta/3$. Notice that if all $\beta_i=1$, this reduces to the result for a Lifshitz metric with hyperscaling violation.

By now, the IR fate of certain spacetimes with intermediate Lifshitz scaling and/or hyperscaling-violation regions has been explored and found to have a re-emergent AdS$_2$ geometry \cite{resolving}. Though the Bianchi-type metrics do not have the rotational invariance of an $AdS_2\times R^3$ geometry, we still expect that they may appear in only a finite range of scales, and may cross over to some other fixed point for $r\gg r_F$. Thus, the nature of the tidal forces may change before they become singular. In any case, given the fact that the Bianchi-type metrics are not vacuum solutions to the Einstein equations, the nontrivial stress-energy sources will affect propagation of test objects and may resolve any potential singularities due to divergent tidal forces in a manner akin to \cite{bdhs}. Finally, we should point out that slightly heating up the geometry can mask these singular effects.

\section{Basic properties of Heisenberg spacetimes}\label{sec:heis}
Before we consider holography for spaces with Heisenberg isometries, we will introduce the Heisenberg algebra and its associated Lie manifold. We will see that the Heisenberg group has a fundamental connection to the quantum harmonic oscillator, gauge theories with Landau levels, and modular forms

\subsection{The Heisenberg algebra}
The algebra of type II is the Lie algebra of the continuous Heisenberg group. This group has three generators which obey the algebra:
\be
[\xi_x,\xi_z]=\xi_y,~~[\xi_x,\xi_y]=[\xi_z,\xi_y]=0.
\ee
This algebra gets its name because it comes from the Heisenberg uncertainty principle. The fact that position and momentum are non-commuting operators comes from the fact that  a variable and its derivative operator don't commute, i.e. $[x,\partial_x]=1$, and that in the $x$ basis, momentum is $i\hbar\partial_x$. $x,\partial_x$, and the identity form a representation of the type II algebra, where the identity is added so the algebra closes.
A one parameter family of bases for the $\xi_i$ is
\be
\xi_x=\partial_x+(\cos^2\theta) z\partial_y,~~\xi_z=\partial_z-(\sin^2\theta ) x\partial_y,~~\xi_y=\partial_y,
\ee
where $\theta$ is constant. Translation along the $y$-direction is always an isometry, and there is additional translation invariance in the $x-z$ plane in the direction $\sin^2\theta\hat x +\cos^2\theta\hat z$. Note that this has broken both translational and rotational invariance in the $x-z$ plane to translations along a single direction. Choosing $\theta$ will set a gauge in the metric which is invariant under these Killing isometries; we will come back to this point later on when we consider what quantities are gauge invariant.

The vector fields which commute with the above generators are
\be
\partial_y,~~\partial_x-(\sin^2\theta) z\partial_y,~~\partial_z+(\cos^2\theta ) x\partial_y
\ee
and the one forms dual to these vector fields are
\be
dx, ~~dz, ~~dy+(\sin^2\theta) zdx-(\cos^2\theta) xdz.
\ee
The convention we will be using for most of the following computations is $\theta=0$, for which the last 1-form becomes $dy-xdz$. Note that in doing this we have picked a particular gauge which has translation invariance along the $z$-direction.

\subsection{Representations of the Heisenberg group}\label{sec:reps}
The three dimensional Heisenberg algebra is the Lie algebra of the three dimensional continuous Heisenberg group,  $H_3$.
We can represent this group by the set of $3\times 3$ real upper triangular matrices with ones along the diagonal, 
\[ g_i=
\begin{pmatrix}
~1~&~a_i~&~c_i~\\
~0~&~1~&~b_i~\\
~0~&~0~&~1~
\end{pmatrix}
\]
with $a_i,b_i,c_i,\in \mathbb{R},\mathbb{Z}$. Multiplying two elements yields another group element,
\[ g_1g_2=
\begin{pmatrix}
~1~&~a_1+a_2~&~c_1+c_2+a_1b_2~\\
~0~&~1~&~b_1+b_2~\\
~0~&~0~&~1~
\end{pmatrix},
\]
however the commutator of two group elements,
\[ [g_1,g_2]=
\begin{pmatrix}
~0~&~0~&~a_1b_2-a_2b_1~\\
~0~&~0~&~0~\\
~0~&~0~&~0~
\end{pmatrix},
\]
is nonzero; the group is nonabelian with center
\[ \begin{pmatrix}
~1~&~0~&~1~\\
~0~&~1~&~0~\\
~0~&~0~&~1~
\end{pmatrix}.
\]
In the following we will discuss two particular representations of $H_3$, the Schr\"odinger representation and the theta representation.
\subsubsection{The Schr\"{o}dinger representation}
This group has a natural unitary action on the Hilbert space of one-dimensional square integrable functions, $f(x)\in L^2(\mathbb{R})$, of the form,
\be
\pi (g_i)f(x)=c_ie^{\pi i(2xb_i+a_ib_i)}f(x+a_i),
\ee
where $\pi(g_i)$ is the operator  corresponding to the group element $g_i$. Composition of operators obeys
\be
\pi(g_1)\pi(g_2)f(x)=c_1c_2e^{\pi i (a_2b_1-b_2a_1)}e^{\pi i (2x(b_1+b_2)+(a_1+a_2)(b_1+b_2))}f(x+a_1+a_2).
\ee
\subsubsection{The theta representation}
First, consider a space of holomorphic function $f(z):\mathbb C\to \mathbb C$, along with a fixed complex number $\tau\in\mathbb H$, and consider two operators, $S_b, T_a$ which act on $f(z)$ as
\be
(S_b f)(z)=e^{b\partial_z}f(z)=f(z+b)
\ee
and
\be
(T_a f)(z)=e^{2\pi iaz+a\tau\partial_z}f(z)=e^{\pi ia^2\tau+2\pi iaz}f(z+a\tau),
\ee
with $a,b\in \mathbb R$. These operators individually obey a law of composition, $S_{b_1}\circ S_{b_2}=S_{b_1+b_2}, T_{a_1}\circ T_{a_2}=T_{a_1+a_2}$, but they don't commute:
\be
S_b\circ T_a=e^{2\pi iab}T_a\circ S_b.
\ee
Therefore, if we include the action of a unitary phase $\lambda$ along with $S_b, T_a$, we can define a set of unitary operators, $U_\tau(\lambda,a,b)$, for fixed $\tau$ which act on holomorphic functions as
\be
(U_\tau f)(z)=\lambda e^{\pi ia^2\tau+2\pi i az}f(z+a\tau+b)
\ee
which form a representation of $H_3(\mathbb R)$. Note that each $\tau$ specifies a different representation. We can consider $U_\tau$ as a unitary operator on a Hilbert space $\mathcal H_\tau$ of entire functions on $\mathbb C$ with finite norm, where the norm is defined as
\be
||f||_\tau^2=\int_{\mathbb C}dx dy ~e^{-\frac{2\pi (x^2+y^2)}{\Im \tau}}|f(x+iy)|^2,
\ee
where $\Im \tau$ is the imaginary part of $\tau$. In fact, from the action of $U_\tau$ we see that the modular group $SL_2(\mathbb R)$ acting on $\tau$ is an automorphism on this space of representations of $H_3(\mathbb R)$!

Finally, if we define the subgroup $\Gamma_\tau=U_\tau(1,a,b)\in H_3(\mathbb Z)$, we see that the Jacobi theta function, defined as
\be
\theta_3(z; \tau)=\sum_n e^{\pi in^2\tau+2\pi inz},
\ee
has the properties $\theta_3(z+1; \tau)=\theta_3(z;\tau)$ and $\theta_3(z+a\tau+b; \tau)=e^{-\pi ia^2\tau-2\pi iaz}\theta_3(z;\tau)$, and is invariant under the action of $\Gamma_\tau$. In fact, we will see the theta representation crop up again in \S~\ref{sec:scaling}.

\subsection{Particle dynamics in Heisenberg metrics}
Before we investigate the physics of Heisenberg branes from an AdS/CFT point of view, let's analyze the motion of particles in 4-dimensional metrics which have these symmetries. This will correspond to a spacetime slice at constant radius at the scale $r\approx r_F$. In particular, the geodesics will not be the familiar straight lines of four-dimensional Minkowski space because of the spatial anisotropy, but because there exist three spatial Killing vectors for each Bianchi class, there will be three conserved momenta along a geodesic. Calculating the effect of these isometries on motion of test particles in these metrics will be useful when searching for natural conserved quantities in field theory duals once we add back in radial scaling. We can compute geodesic motion in a metric of the form
\be
ds^2=-dt^2+g_{ij}(x)dx_idx_j
\ee
by computing Euler-Lagrange equations from the Lagrangian $L=\frac{1}{2}g_{\mu\nu}\dot x^\mu\dot x^\nu$ (where $\dot x^\mu$ is a derivative with respect to an affine parameter $\lambda$) and imposing the constraint equation $2L=\epsilon$, where $\epsilon=-1$ for timelike geodesics and $\epsilon=0$ for null geodesics. Since the metric for each of these Lagrangians is manifestly independent of $t,y,z$, there will be a conserved energy $E$ and two conserved momenta, $p_y, p_z$ found from varying $\frac{\partial L}{\partial \dot y}$ and $\frac{\partial L}{\partial \dot z}$. However, we know there is one additional Killing vector of the metric, so this will give us an additional first integral of motion along the geodesic. In this section, we will compute geodesics in these backgrounds and find the generalized conserved momenta.

On a constant $r$ slice, the metric will be of the form
\be
ds^2=-dt^2+dx^2+(dy-xdz)^2+dz^2,
\ee
and the Lagrangian will be
\be
\mathcal L=\frac{1}{2}(-\dot t^2+\dot x^2 +(\dot y-x\dot z)^2+\dot z^2).
\ee
The three Killing vectors, $\partial_t,\partial_y,\partial_z$ lead to the conserved quantities: $E=-\dot t, p_y=\dot y-x\dot z$, and  $p_z=-x\dot y +(1+x^2)\dot z$.  The equation of motion is
\be
\ddot x=-\dot z(\dot y-x\dot z)=-p_y(p_z+xp_y),
\ee
and the constraint equation is
\be
\dot x^2+(p_z+xp_y)^2-E^2+p_y^2=\epsilon.
\ee
Noting that $\dot y-x\dot z=p_y$, we can write the equation of motion as a total derivative,
\be
\frac{\partial}{\partial \lambda}\left (\dot x+zp_y\right )=0,
\ee
so we see that there is yet another conserved momentum along the geodesic, $ p_x=\dot x+zp_y$ which is conjugate to the Killing vector $\partial_x+z\partial_y$. Rewriting the constraint equation in terms of all four conserved quantities, we get that the surface of a geodesic satisfies
\be
(x^2+z^2)p_y^2+2p_y(xp_z-z p_x)+\sum_i \vec p_i^2-E^2-\epsilon=0.
\ee
At fixed energy and momenta, this is the equation of an circle in the $x-z$ plane, and in three dimensions it is a spiral with height proportional to the arc length of the circle, $y(\lambda)\sim\int x(\lambda)dz$. This is the geodesic for the motion of an electrically charged particle in a constant magnetic field which is pointing in the $\hat y$-direction.

\section{Addition of (hyper)scaling (violation)}\label{sec:scaling}
Now that we are familiar with the Heisenberg algebra, its representations and metric properties, we will investigate the implications of Heisenberg isometries in holography. As discussed in \S~\ref{sec:rev}, when we include a radial direction, there is freedom to add a general power of $r$ in front of each one-form invariant under the Heisenberg symmetries. Therefore, the most general near-horizon metric of an extremal black brane with the symmetries of the type II algebra is
\be\label{eq:ds2}
ds^2=L^2r^{2\theta/ 3}\left (-r^{-2\beta_t}dt^2+{dr^2\over r^2}+r^{-2\beta_x}dx^2+r^{-2\beta_z}dz^2+r^{-2(\beta_x+\beta_z)}(dy-xdz)^2\right ),
\ee
where we have chosen the $\eta_{ij}$ of equation (\ref{eq:gen_metric}) to be the identity matrix, and we have included the possibility of a hyperscaling violation exponent, $\theta$. By writing down a metric (based on a particular basis we have chosen for the Heisenberg algebra,) we are implicitly picking a bulk gauge. However, physical quantities in the dual field theory should be independent of the gauge we choose in the bulk. We will see this explicitly when we two point functions for probe scalar field in \S~\ref{sec:2pt}.

 The NEC which govern the physically allowed values of the parameters $\{\beta_i\}, \theta$ are given in full in appendix \ref{sec:NEC}. Here we will list a few particular cases. 
\begin{itemize}
\item $\beta_t=1,\theta=0$

This theory is scale-invariant but not Lorentz-invariant. As was the case in \S~\ref{sec:t1NEC}, the constraints are symmetric in $\beta_x,\beta_z$, so we get that if $\beta_x=0$, then $0\leq \beta_z\leq\frac{1+\sqrt 5}{4}$, and vice versa. If $\beta_x=\beta_z=\beta$, then the allowed range for $\beta$ is $0\leq\beta\leq 2/3$. We will solve for scalar field dynamics in theories of this type in \S~\ref{sec:2pt}.
\item $\beta_x=\beta_z=0$

In this case the three spatial directions don't scale, but there is a scaling symmetry in time and an overall hyperscaling-violation exponent. The constraints are $\theta^2-3\beta_t\theta\geq 0$ and $\beta_t(\beta_t-\theta)\geq 1/2$ which implies that if $\theta=0$ then $\beta_t^2\geq 1/2$.
\item $\beta_t=1,\beta_z=0$

The constraints in this case reduce to $\beta_x\geq {\theta\over 2}-{1\over 4}$, $\beta_x(1-\beta_x)\geq {\theta\over 2}-{\theta^2\over 6}$, and $\beta_x(1+\theta-2\beta_x)\geq -{1\over 2}$.
\end{itemize}
As shown in \cite{bigteam}, for $\theta=0$ this system can be supported by an action of the form of equation (\ref{eq:gen_action}) where the gauge field has the form $A_t=A_0r^{-\beta_t}$. If we set $L=1$ then the solutions of the equations of motion leave one free parameter. In terms of $\beta_t$, the other parameters are
\bea\nonumber
&&m^2=-\beta_t\left(\beta_t+\sqrt{16+\beta_t^2}\right),~\Lambda=-\frac{1}{32}\left (72+19\beta_t^2+3\beta_t\sqrt{16+\beta_t^2}\right )\\
&&A_0^2=\frac{19}{8}+\frac{3\sqrt{16+\beta_t^2}}{8\beta_t},~\beta_x=\beta_z=-{1\over 8}\left (\beta_t+\sqrt{16+\beta_t^2}\right ).
\eea
Note that this is slightly different from the solution written in \cite{bigteam} as our conventions for the action and the metric differ.

We should also note that it has not yet been shown that five-dimensional type II spacetimes can be embedded into asymptotic $AdS_5$, though this is the assumption under which we have been working.\footnote{There is work underway which will show that there are such metrics one can write down which obey reasonable energy conditions \cite{bigteam3}.} We also will not do a stability analysis of the solutions we present in this section. Such an analysis would allow us to map out phase diagrams involving these metrics in more detail. 

In the rest of this section we discuss five-dimensional hyperscaling violating type II solutions (\S~\ref{sec:5hyp}), four-dimensional hyperscaling violating solutions from dimensional reduction of five-dimensional type II (\S~\ref{sec:KK}), and RG flows between solutions with type II isometries (\S~\ref{sec:RG}). In \S~\ref{sec:type2geodesic} we compute two-point functions for massive scalars in the semiclassical limit, and we briefly discuss entanglement in \S~\ref{sec:EE}.
\subsection{Five-dimensional hyperscaling violation}\label{sec:5hyp}

In this section we will look for an action which supports a metric of the form of equation (\ref{eq:ds2}) with $\theta\neq 0$. This was done for a few of the other Bianchi types in \cite{bigteam2}, and we find the solution for type II is quite similar. The metric is scale invariant up to an overall power, i.e. if we rescale $x_i\to \lambda^{\beta_{x_i}}x_i$, $r\to \lambda r$, then the metric transforms as $ds\to \lambda^{\theta/3}ds$. We will now set $L=1$ for the rest of this section.

We can support a metric of this form by considering the action,
\be
S=\int d^5 x\sqrt{-g}\left \{R-2\Lambda e^{2\delta\phi}-\frac{e^{2\alpha\phi}}{4}F^2-\frac{1}{2}(\nabla\phi)^2-\frac{m^2}{4}e^{2\epsilon\phi}A^2\right \},
\ee
where the dilaton takes the form $\phi=k\log r$ and the gauge field has the form $A_t=A_0r^\gamma$. The stress-energy tensor for this action is
\be
T_{\mu\nu}=-{g_{\mu\nu}\over 2}\mathcal L+\frac{e^{2\alpha\phi}}{2}F_{\mu\rho}F_{\nu\sigma}g^{\rho\sigma}+\frac{m^2}{4}e^{2\epsilon\phi}A_\mu A_\nu+\frac{1}{2}\partial_\mu\phi\partial_\nu\phi,
\ee
which enters into the Einstein equations as $R_{\mu\nu}-{1\over 2}Rg_{\mu\nu}=T_{\mu\nu}$.
The Einstein equations, scalar and gauge field equations of motion become algebraic if we take
\be
\epsilon=\alpha+\delta; ~~\theta+3\delta k=0; ~~\beta_x=\beta_z=\beta; ~~\gamma+\beta_t+(\alpha+\delta)k=0.
\ee
Using these relations, the scalar equation of motion reduces to
\be
k(\gamma+\alpha k-2\delta k-4\beta)=4\delta\Lambda-\alpha\gamma^2A_0^2-\frac{m^2(\alpha+\delta)A_0^2}{2}
\ee
and the gauge field equation of motion becomes
\be
\frac{m^2}{2}=\gamma(\alpha k-2\delta k-4\beta).
\ee
Finally, we have to consider the non-redundant Einstein equations, which are
\bea\nonumber
&&{1\over 4}+11\beta^2+{1\over 8}A_0^2(m^2+2\gamma^2)+12\beta\delta k+{k^2\over 4}+3k^2\delta^2+\Lambda=0\\\nonumber
&&{1\over 4}+5\beta^2-{1\over 8}A_0^2(m^2-2\gamma^2)-4\beta\gamma-4\beta k(\alpha-2\delta)-{k^2\over 4}-3k(k\alpha+\gamma)\delta+3k^2\delta^2+\Lambda=0\\\nonumber
&&-{1\over 4}+7\beta^2-{1\over 8}A_0^2(m^2+2\gamma^2)+\gamma^2-3\beta(\gamma+k(\alpha-2\delta))+k\gamma(2\alpha-\delta)+{k^2\over 4}+k^2(\alpha^2-\alpha\delta+\delta^2)+\Lambda=0\\\nonumber
&&{3\over 4}+3\beta^2-{1\over 8}A_0^2(m^2+2\gamma^2)+\gamma^2-2\beta(\gamma+k(\alpha-2\delta))+k\gamma(2\alpha-\delta)+{k^2\over 4}+k^2(\alpha^2-\alpha\delta+\delta^2)+\Lambda=0.
\eea
These equations have 3 free parameters. Defining
\be
C_1={1\over \beta(1+6\alpha\delta)}; ~~C_2=\beta^2(4+11\alpha^2+4\alpha\delta-4\delta^2),
\ee
we can obtain a solution if the following equations are satisfied,
\bea\nonumber
k&=&C_1(2\alpha-11\beta^2\alpha+2\delta-2\beta^2\delta)\\\nonumber
\beta_t&=&C_1(1-6\delta^2+\beta^2(-4+9\alpha\delta+6\delta^2))\\\nonumber
m^2&=&2C_1^2(1+2\alpha^2+4\delta(\alpha-\delta)-C_2)(-2(\alpha-2\delta)(\alpha+\delta)+C_2)\\\nonumber
\gamma&=&C_1(C_2+4\delta(\delta-\alpha)-2\alpha^2-1)\\\nonumber
A_0^2&=&\frac{2-11\beta^2+12\delta^2(\beta^2-1)}{1+2\alpha^2+4\delta(\alpha-\delta)-C_2}\\\nonumber
\Lambda&=&{C_1^2\over 4}(-2-4\alpha^2+\beta^2(12\delta^2-11)C_2-4\delta(5\alpha-2\delta+12\alpha^2\delta+6\alpha\delta^2+12\delta^3)+\\&&2\beta^2(5-50\delta^2+48\delta^4+\alpha^2(22-42\delta^2)+\alpha\delta(5-12\delta^2)))
\eea
with $\beta,\alpha,\delta$ free. We will get constraints on the allowed values and parameter combinations from the fact that we require $\beta_t, \beta >0$, $A_0^2>0$, and $\Lambda <0$.

To convince yourself that this solution makes sense for reasonable numerical values of  the parameters, here we quote one particular numerical solution. One choice of metric parameters that satisfies the NEC is $\theta=1,\beta_t=1,\beta={1\over 2}$. In this case we can get a solution if we set, e.g., $k=1, \delta=-{1\over 3}, \alpha=4$, which makes $C_1=-{2\over 7}$ and $C_2=44-{4\over 9}$. Plugging these in, the rest of the parameters become
\be
m^2=-{8\cdot 28\over 9},~\gamma=-{14\over 3},~A_0^2={3\over 28}, ~\Lambda=-{22\over 3},
\ee
which seems reasonable. Note that $m^2<0$, however this is allowed as the BF bound for systems with a hyperscaling violation exponent allows for more negative masses than in pure AdS space. We also have that $\gamma <0$, though this is also what is found in \cite{bigteam2}.

\subsection{Kaluza-Klein reduction to four dimensions}\label{sec:KK}
We can also obtain hyperscaling violating solutions in four dimensions by considering KK reduction along the $y$-direction.  When we compactify a single direction, we get two additional fields in the lower-dimensional action, a massless gauge field arising from the $g_{yi}$ metric modes and a scalar field which comes from the $g_{yy}$ metric mode. We will see below that the metric gauge field will become a bulk magnetic field in four-dimensions and the scalar mode will give rise to a hyperscaling violation exponent. We follow the method of \cite{bigteam2}, which has a nice review of how KK reduction works for five-dimensional Bianchi type metrics.

When considering dimensional reduction along the $y$-direction, we can write the five-dimensional type II metric, $ds_5^2$, in terms of the dimensionally reduced four-dimensional metric $ds_4^2$ as
\be
ds_5^2=e^{2\alpha\phi}ds_4^2+e^{2\gamma \phi}(dy+B_\mu dx^\mu)^2,
\ee 
where $B_\mu$ is is a gauge field along the compactified direction. We also write the five-dimensional massive vector field in terms of the four-dimensional one as
\be
A^5=A_\mu^4dx^\mu+A_ydy,
\ee
which tells us immediately that $A_\mu^4=(A_0r^{\beta_t},0,0,0)$.
The action supporting this metric will have the form
\be
S=\int d^4x\sqrt{-g_4}\left (R_4-2\Lambda e^{2\alpha\phi}-{1\over2}\partial_\mu\phi\partial^\mu\phi-{1\over 4}e^{-6\alpha\phi}H^2-{1\over 4}e^{-2\alpha\phi}(F_4)^2-{1\over 4}m^2A_\mu^4(A^4)^\mu\right )
\ee
where $H=dB$ is the field strength of the gauge field which comes from the dimensional reduction.

From the metric of equation (\ref{eq:ds2}), it is easy to read off the solution for the fields in four dimensions. $B_\mu=(0,0,0,-x)$, and the scalar field solution is
\be
\phi(r)=-\frac{\beta_x+\beta_z}{\gamma}\log r,
\ee
and the 4-dimensional metric is
\be
ds_4^2=r^{\frac{2\alpha}{\gamma}(\beta_x+\beta_z)}\left (-r^{2\beta_t}dt^2+\frac{dr^2}{r^2}+r^{-2\beta_x}dx^2+r^{-2\beta_z}dz^2\right ).
\ee
This is just a four-dimensional, translationally invariant, type I hyperscaling violating solution. By dimensionally reducing along the $y$-direction, our solution is no longer anisotropic! We see that in addition to the massive gauge field, there is a constant magnetic field coming from the new $B_\mu$ gauge field, and a log-running scalar supporting the hyperscaling-violating geometry.
Also, when we require the four-dimensional action to take the conventional Einstein-Hilbert form with canonically-normalized dilaton kinetic term, this requires $\alpha=-1/2\sqrt{3}$ and $\gamma=1/\sqrt{3}$. Thus, we see that the hyperscaling violation exponent is $\theta=-(\beta_x+\beta_z)$. Typically, $\beta_x$ and $\beta_z$ are both $\geq 0$; this leads to generic values of $\theta$ which are negative or zero. Theories with negative $\theta$ are allowed from the NEC, and in fact are known to arise in the near-horizon region of Dp-branes compactified on an $S^{8-p}$ for $p\neq 3$. The latter have a hyperscaling violation exponent given by \cite{theta}
\be
\theta=p-\frac{9-p}{5-p};
\ee
thus we see that for $\beta_x+\beta_z={1\over 3}$, this matches the result for the near-horizon region of a D2-brane with a compact $S^6$. In \S~\ref{sec:2pt}, we will study the cases of $\beta_x=\beta_z=0$ (this is just $AdS_2\times R^2$ after KK reduction) and $\beta_x=\beta_z={1\over 2}$, which leads to $\theta=-1$ in four dimensions.

There is an extensive literature on spacetimes of this form; it is quite interesting that they are related to Heisenberg branes in five dimensions through compactification, as we elaborate on in the next section.

\subsection{RG Flows}\label{sec:RG}

In the five-dimensional metric solutions we have discussed, there are two vector fields: a massive vector field in the action which supports the metric of the form $A_t=A_0r^\gamma$, and a massless gauge field, $B_\mu$ in the metric which generates the anisotropy and which we can think of as a magnetic field. In this section we will consider RG flows that are governed by the behavior of either or both of these vector fields.

\subsubsection{Heisenberg phases from Higgsing}\label{sec:higgsing}

Let us consider how a metric of the form (\ref{eq:ds2}) might arise as an IR phase of some asymptotically AdS geometry. Consider a probe complex scalar field, $\psi$, electrically charged under a massless gauge field. Schematically, this may have the form
\be
S=\int d^5 x \sqrt{-g}\left (R-2\Lambda-\frac{F^2}{4}-|\nabla_\mu\psi|^2-m^2|\psi|^2\right )
\ee
with covariant derivative $\nabla_\mu=\partial_\mu+iA_\mu$ where $A_t=A_0r^\gamma, A_i=0$. We could imagine the background spacetime to be an extremal electrically charged Reissner-Nordstrom black brane, which can be supported by a massless gauge field. This geometry is asymptotically $AdS_5$ and has a near-horizon $AdS_2\times R^3$ region with extensive entropy density. Now consider the backreaction of $\psi$ on the gauge field. If we tune the mass of the scalar so that it becomes unstable to a superconducting transition a la \cite{hhh}, $\psi$ will get a vev $\langle \psi\rangle$. If this vev is non-normalizable, it will generate a mass term for the gauge field from the derivative term in the action: $|\nabla_\mu\psi|\sim A_\mu A^\mu|\langle\psi\rangle|^2$. This leaves an action of the form of equation (\ref{eq:gen_action}), which has the Bianchi type II metric as a solution! This breaks the gauge symmetry. We would have to do further analysis to determine in what parameter range (if any) this solution would be energetically preferable. We should also comment that a similar mechanism to generate a mass for a vector field could lead to phases with the symmetries of many of the other Bianchi types, as they are also supported by massive vector fields.\footnote{This would very likely only be the case in a finite range of radial scales, as the gauge field mass generated by $\psi$ is of order the AdS scale and would not be stable at a fixed value.}

\subsubsection{Instabilities in four dimensions}
In \S~\ref{sec:KK} we saw that compactifying along the direction of the magnetic field in the five-dimensional metric leads to a four-dimensional metric with hyperscaling violation or an $AdS_2\times R^2$.  We know the former solutions do not continue into the deep IR \cite{resolving} and the latter may be unstable to inhomogeneous phases \cite{vlattice}. In the language of \S~\ref{sec:higgsing}, when we compactify, a constant vev for $\psi$ is no longer a solution to the equations of motion because of the constant magnetic field coming from $dB$. We may expect the true ground state to be an inhomogenous phase, where $\psi$ condenses to a vortex lattice \cite{vlattice}. It is perhaps not a surprise that such a lattice solution can be given by a Jacobi-$\theta$ function, which is a natural representation of the Heisenberg algebra (see \S~\ref{sec:reps}). We will see in the next section that the two-point function of a (five-dimensional) probe scalar has the structure of a theory with Landau levels, as one would expect in the four-dimensional theory, where the scalar is charged and coupled to the magnetic field.

\subsection{Scalar correlators in the semiclassical approximation}\label{sec:type2geodesic}
Before we solve the scalar equation of motion in the type II background, we first consider the two-point function for a massive scalar using the geodesic approximation. This analysis will be very similar to that of \S~\ref{sec:typeIgeodesic}.
The action for a massive scalar in the metric of equation (\ref{eq:ds2}) is
\be
S=-m\int d\lambda \sqrt{r^{-2\beta_t}\dot\tau^2+{\dot r^2\over r^2}+r^{-2\beta_x}\dot x^2+r^{-2\beta_z}\dot z^2 + r^{-2(\beta_x+\beta_z)}(\dot y-x\dot z)^2 },
\ee
where once again $\tau=it$ and $\lambda$ is an affine parameter along the geodesic which we will choose to be $r$. (Note that we have set $\theta=0$ and $L=1$.) Because there is manifest translation invariance in the $y,z$ directions, the action will be a function $S(x_i,x_f,|\Delta y|,|\Delta z|)$ where $x_i$ and $x_f$ are the initial and final values of $x$. 
In the following we will consider several cases. 
\begin{enumerate}[(i)]
\item $\Delta z=0$

This case is simplest because it removes the manifest dependence of the action on the $x$ coordinate. This exactly reduces to the results found in \S~\ref{sec:typeIgeodesic}, where the propagator is a function only of $|\Delta x_i|~\forall ~x_i$. We expect scale invariance between points which are only separated along $x, y$ and $t$, independent of the location along the $z$-direction.

\item $\Delta x=0, \Delta \tau=0$

In this case $x$ takes a constant value along the geodesic, $x=x_0$. We can make the coordinate transformation $\tilde y=y-x_0z$ to put the action in the form, 
\be
S=-m\int dr \sqrt{{1\over r^2} + r^{-2(\beta_x+\beta_z)}\tilde{y}'^2 + r^{-2\beta_z}z'^2)}
\ee
which has the same functional form as equation (\ref{eq:typeIaction}) for the action in an anisotropic type I background. If we set $\beta_x=0$ we get a propagator of the form of equation (\ref{eq:powerlawG}),
\be
G(|\Delta y|,|\Delta z|)\sim\left (\frac{ \epsilon}{\sqrt{\Delta \tilde y^2+\Delta z^2}}\right )^{2m/\beta_z}=\left (\frac{ \epsilon}{\sqrt{\Delta y^2-2x_0|\Delta y||\Delta z|+(1+x_0^2)\Delta z^2}}\right )^{2m/\beta_z},
\ee
where $x_0$ is a constant. For $x_0\ll\frac{\Delta y^2+\Delta z^2}{2|\Delta y||\Delta z|}$, this has the expansion
\be
G(|\Delta y|,|\Delta z|)\sim\left (\frac{ \epsilon}{\sqrt{\Delta y^2+\Delta z^2}}\right )^{2m/\beta_z}\left (1+\frac{2mx_0}{\beta_z}\frac{|\Delta y||\Delta z|}{\Delta y^2+\Delta z^2}\right ).
\ee
We get power law behavior in the separation distance in the $y-z$ plane at leading order. For $x_0\gg \frac{2|\Delta y|}{|\Delta z|}$ the propagator has the expansion,
\be
G(|\Delta y|,|\Delta z|)\sim\left (\frac{ \epsilon}{x_0|\Delta z|}\right )^{2m/\beta_z}\left (1+\frac{2m}{x_0\beta_z}\frac{|\Delta y|}{|\Delta z|}\right ),
\ee
which goes like a power law in the distance along $z$ and is independent of the separation along $y$ at leading order.

Setting $\beta_z=0$ we get a propagator of the form of equation(\ref{eq:mixedG}),
\be
G(|\Delta y|,|\Delta z|)\sim e^{-m\sqrt{\Delta z^2+\frac{\log^2\left ({\beta_x^2\Delta \tilde y^2\over \epsilon^{2\beta_x}}\right )}{\beta_x^2}}}=e^{-m\sqrt{\Delta z^2+\frac{\log^2\left ({\beta_x^2(|\Delta y|-x_0|\Delta z|)^2\over \epsilon^{2\beta_x}}\right )}{\beta_x^2}}},
\ee
which is of mixed exponential and power law form.

One could also do this case for different values of the critical exponents $\beta_x$ and $\beta_z$; this reduces to the scenario discussed in \S~\ref{sec:typeIscaling}.

\item $\Delta y=0,\Delta \tau=0, \beta_x=\beta_z=0$

In this case we just get the geodesic distance in the $x-z$ plane. We'll approximate the metric as a perturbation of flat space by introducing a small parameter, $\epsilon\ll 1$.  We are free to do this, as this amounts to rescaling $x$ in the one-form, $dr-xdz\to dr-\epsilon xdz$, which appears in the metric. 

Now the action depends on $x$ but not $r$ as,
\be
S=-m\int d\lambda \sqrt{ x'^2 +  (1+ \epsilon^2 x^2)z'^2}.
\ee
Taking $\lambda=x$, there is a conserved momentum conjugate to  $z$. Going through the same process as above, and keeping only terms of zeroth order and $O(\epsilon^2)$, the action along the geodesic evaluates to 
\be\label{eq:anisotropicaction}
S\sim-m\sqrt{\Delta x^2+\Delta z^2}\left (1+\frac{\epsilon^2(x_f^3-x_i^3)}{3\Delta x^3}(\Delta x^2+\Delta z^2)\right )
\ee
which reduces to the exponential form when $\epsilon\to 0$, as expected. At $O(\epsilon^2)$, the propagator now has a dependence on $x_f^3-x_i^3$ due to the explicit $x$-dependence of the metric.
\end{enumerate}

Though we have only listed a few special cases which we found analytically tractable, one could of course compute semiclassical geodesics numerically  for more general critical exponents and nonzero separation in $\{\Delta\tau,\Delta x,\Delta y,\Delta z\}$. Once again, as discussed in \S~\ref{sec:typeIscaling}, the approximation will only be valid in the IR regime $m|\Delta \vec x|\gg 1$ in units of the scale $r_F$.

\subsection{Entanglement entropy}\label{sec:EE}

Finally, we consider entanglement entropy for a strip on the boundary in the five-dimensional type II metric of equation (\ref{eq:ds2}), where we set $\theta=0$. The form of this computation is very similar to the computation of massive geodesics in the previous section. The case which is immediately tractable is for a strip of length $-{L_{y,z}\over 2}\leq y,z\leq {L_{y,z}\over 2}$ and width $-{\ell \over 2}\leq x\leq {\ell \over 2}$, with $\ell\ll L_{y,z}$. In this case, since the metric only has explicit dependence on $x,r$, the radial profile will be independent of $y,z$, and the area of the minimal surface will be given by the action,
\be
\mathcal A\sim L_yL_z\int_{r_t}^\epsilon dr~ r^{-1-3\beta}\sqrt{1+ r^{2-2\beta}x'^2}
\ee
where we have set $\beta_x=\beta_z=\beta$, and $r_t$ is the turning point in the bulk. This is for $\beta\neq 0$. At $\beta=0$, the spatial directions decouple and the minimal surface does not travel into the bulk.

We get a set of equations for $\ell$ and $\mathcal A$ as a function of $r_t$ that can be solved numerically for general values of $\beta$:
\be\label{eq:EE1}
\ell = \int_{r_t}^0 dr {r^{\beta-1} r_t^{-4\beta}\over\sqrt{r^{-8\beta}-r_t^{-8\beta}}},
\ee
and
\be\label{eq:EE2}
\mathcal A= 2L_yL_z\int _{r_t}^\epsilon dr {r^{-1-7\beta}\over\sqrt{r^{-8\beta}-r_t^{-8\beta}}}.
\ee

In four dimensions, the boundary between entanglement regions in the field theory is a one-dimensional surface, while minimal surface in the bulk is two-dimensional. From the four-dimensional perspective, the five-dimensional computation of entanglement entropy gives a minimal entangling surface which wraps the compact $y$ direction. We see the effects of this compact dimension in the formulas for the minimal surface, equations (\ref{eq:EE1}) and (\ref{eq:EE2}), as there are powers of the radius which come from the $g_{yy}$ metric factor. Thus, we we can view this computation as giving the entanglement for the KK-reduced theory of \S~\ref{sec:KK}, where now $L_y$ is the size of the compact circle.

If we rotate the strip so that  $-{L\over 2}\leq x,y\leq {L\over 2}$ and  $-{\ell \over 2}\leq z\leq {\ell \over 2}$, with $\ell\ll L$, the profile of the minimal surface will have dependence on both $x$ and $z$, since the metric has explicit dependence on $x$. However, we expect the entanglement entropy of the field theory should be independent of the gauge we choose for the bulk metric, so we should ultimately get a similar result to what is above.

\section{Two point functions of scalar probes in type II}\label{sec:2pt}
In this section we investigate the dynamics of probe scalar fields in scale-invariant type II spacetimes.  We generalize the familiar AdS/CFT prescription \cite{Witten1,GKP} for computing boundary correlators to spacetime-dependent metrics in \S~\ref{sec:st2pt}, and we apply this to both massless and massive scalar probes in \S~\ref{sec:LL2pt}. The dynamics of a charged particle in a constant magnetic field will be evident in the scalar two-point function which organizes itself into Landau levels, with additional behavior due to scaling in certain coordinate directions. We also discuss the interpretation of these scalar two-point functions in the four (bulk) dimensional KK-reduced geometry with magnetic field. We find that by doing the calculation using the five-dimensional type-II metric, we can see the effect of the 4D magnetic field on the 4D charged scalars which result from KK-reduction of the probe 5D scalar, thus allowing us to go beyond the four-dimensional probe approximation.

\subsection{Computing a scalar Green's function in a spacetime-dependent metric}\label{sec:st2pt}

In this section we will review the familiar prescription for computing two-point functions using the AdS/CFT dictionary a la \cite{Witten1,GKP}, generalizing to a metric which now depends on one of the spacetime coordinates, $g_{\mu\nu}=g_{\mu\nu}(r,x)$. In particular, here we define what we mean by a ``momentum space" two-point function, $\langle\mathcal O(x,\vec k)\mathcal O(x',-\vec k)\rangle$, and a ``position space" two-point function, $\langle\mathcal O(t,\vec x)\mathcal O(t',\vec x')\rangle$, both of which we will compute in the next section.

Here we briefly repeat a few basics mentioned in \S~\ref{sec:typeIscaling}. To compute the two-point function of a scalar operator in the dual field theory, we consider a  scalar field $\phi(r,t,\vec x)$ in the bulk with action
\be\label{eq:BULKaction}
S=\int d^5x\sqrt{-g}(-(\partial_\mu \phi)^2-m^2\phi^2)
\ee
which couples to a scalar operator $\mathcal O(t,\vec x)$ in the dual field theory through a term
\be
\int d^4x\phi_0(t,\vec x)\mathcal O(t,\vec x)
\ee
on the boundary. The $\langle \mathcal O\mathcal O\rangle$ two-point function in the field theory is then given by the second derivative of the action with respect to the boundary value of the bulk field,
\be
\langle \mathcal O(t,\vec x)\mathcal O(t',\vec x')\rangle=\frac{\delta^2}{\delta \phi_0(t,\vec x)\delta \phi_0(t',\vec x')}S|_{r\to \epsilon},
\ee
where we require $\phi$ to satisfy the bulk on shell action, with equation of motion
\be\label{eq:phieom}
\partial_\mu(\sqrt{-g}g^{\mu\nu}\partial_\nu\phi(r,t,\vec{x}))=\sqrt{-g}m^2\phi(r,t,\vec x)
\ee
with the boundary condition
\be\label{eq:phibc}
\phi(r,t,\vec x)|_{r\to \epsilon}=\phi_0(t,\vec x).
\ee
As was made clear in \cite{Witten1}, one way to write $\phi(r,t,\vec x)$ such that it satisfies the boundary condition of equation (\ref{eq:phibc}) is in terms of the boundary value $\phi_0$ integrated against a Green's function which becomes a delta function source at the boundary, $r\to\epsilon$:
\be\label{eq:Greensf}
\phi(r,t,\vec x)=\int d^4 x'\phi_0(t',\vec x')G(r,t,\vec x;\epsilon,t',\vec x'),
\ee
where $G$ satisfies equation (\ref{eq:phieom}) with boundary condition $G|_{r\to 0}=\delta(t-t')\delta^{(3)}(\vec x-\vec x')$. 

Our main task in the next section will be computing this Green's function in a type II background. This form of metric has components which depend on both $r$ and $x$; however, it will be useful to Fourier transform in the translation-invariant directions. Thus we will be looking for Green's functions of the form
\be\label{eq:Gexpansion}
G(r,t,\vec x;\epsilon,t',\vec x')=\int d^3\vec ke^{i\omega(t-t')+ik_y(y-y')+ik_z(z-z')}g(r,\vec k,x,x'),
\ee
where $\vec k=(\omega,k_y,k_z)$, and the boundary condition for the momentum-space Green's function is $g(r,\vec k,x,x')|_{r\to \epsilon}=\delta(x-x')$.
To summarize, the Fourier transformed version of equation (\ref{eq:Greensf}) becomes
\be\label{eq:FTgreen}
\tilde\phi(r,x,\vec k)=\int dx'\tilde\phi_0(x',\vec k)g(r,\vec k,x,x').
\ee

Once we have a solution for $\phi$ in terms of its boundary value and the Green's function, we need to plug this back into the action and differentiate with respect to $\phi_0$ to compute the two-point function. First, integrating the bulk action of equation (\ref{eq:BULKaction}) by parts, we get that
\be\label{eq:BULKaction2}
S=\int d^5x~ \left (-\partial_\mu(\phi\sqrt{-g}g^{\mu\nu}\partial_\nu\phi)+\phi(\partial_\mu(\sqrt{-g}g^{\mu\nu}\partial_\nu\phi)-\sqrt{-g}m^2\phi)\right ).
\ee
The second term is just the equations of motion in the bulk, which evaluate to zero, and the first term is a boundary term which we require to evaluate to zero in the $t,\vec x$ directions in order for the action to be finite. This equates to the boundary condition on the Green's function $G|_{x\to\pm\infty}=0$ (in fact, we will have the stronger requirement that the $x$-dependent piece of $G$ to be square-integrable.) When this is satisfied, plugging equation (\ref{eq:FTgreen}) into (\ref{eq:BULKaction2}), the bulk action reduces to\footnote{We have been a bit lax in our notation; the $\vec x$ in equations (\ref{eq:fullaction}) and (\ref{eq:full2ptfn}) runs over the translation invariant directions $(t,y,z)$ only, and does not, in fact, include the $x$ variable.}
\be\label{eq:fullaction}
S=\int d^3\vec kd^3\vec q\int dxdx' e^{i\vec k\cdot \vec x+i\vec q\cdot\vec x'}\tilde\phi_0(x,\vec k)\tilde\phi_0(x',\vec q)\delta(\vec k+\vec q)\sqrt{-g}g^{rr}\partial_rg(r,q,x,x')|_{r\to 0}
\ee
where we have used the facts that $g(r\to\infty)=0$ and $g(r\to\epsilon)=\delta(x-x')$. Differentiating twice with respect to $\phi_0$, we get that the full two-point function is 
\bea\label{eq:full2ptfn}\nonumber
\langle\mathcal O(x)\mathcal O(x')\rangle&=&\int d^3\vec kd^3\vec q e^{i\vec k\cdot \vec x+i\vec q\cdot\vec x'}\delta(\vec k+\vec q)\langle\mathcal O(x,\vec k)O(x',\vec q)\rangle\\
&=&\int d^3\vec k e^{i\vec k\cdot (\vec x-\vec x')}\sqrt{-g}g^{rr}\partial_r g(r,-\vec k,x,x')|_{r\to \epsilon}
\eea
where $\langle\mathcal O(x,\vec k)\mathcal O(x',\vec q)\rangle$ is the two point function in the momentum space of the translation-invariant directions. In the next section we apply the above method to a probe scalar in a five-dimensional metric with Heisenberg symmetry.

\subsection{Probe scalar Green's function in five-dimensional type II}\label{sec:LL2pt}
Now we will use the method outlined in the previous section to compute the two-point function for a probe scalar in a scale invariant type II background (i.e. $\theta=0$.) We begin by solving the equation of motion for a massive scalar with action (\ref{eq:BULKaction}) in a background with metric (\ref{eq:ds2}), which is,
\be
\left(r^{\alpha+2\beta_t}\partial_\tau^2+\partial_r(r^{\alpha+2}\partial_r)+r^{\alpha+2\beta_x}\partial_{x}^2+r^{\alpha+2\beta_z}((x^2+r^{2\beta_x})\partial_{y}^2+\partial_{z}^2+2 x\partial_{y}\partial_{z})-r^\alpha m^2\right)\phi(\tau,r,\vec{x})=0,
\ee
where $\alpha=-2(\beta_x+\beta_z)-\beta_t-1$ and we have transformed to Euclidean time, $\tau=it$. As described in \S~\ref{sec:st2pt}, we Fourier transform the equation for the bulk field in the translational-invariant directions to get an equation of motion for the Green's function $g(r,\vec k,x,x')$ of equation (\ref{eq:Gexpansion}),
\be\label{eq:gauge_green}
\left(-r^{2\beta_t}\omega^2+r^2\partial_r^2+(\alpha +2)r\partial_r+r^{2\beta_x}\partial_{x}^2-r^{2(\beta_x+\beta_z)}k_y^2-r^{2\beta_z}(k_z+xk_y)^2-m^2\right)g(r,\vec k,x,x')=0,
\ee
where we have also divided by $r^\alpha$. The dependence of $g$ on $x'$ will enter through boundary conditions. 

If we take $\beta_x=\beta_z=\beta$, we can assume a separable solution of the form $g(r,\vec k,x,x')=R(r)X(x)$; we will continue with this case for the rest of this section. Once we separate variables we get two equations:
\be\label{eq:xeq}
X''(x)-(k_yx+k_z)^2X(x)=-\lambda X(x)
\ee
and
\be\label{eq:req}
r^2R''(r)+(\alpha+2)rR'(r)-(r^{2\beta_t}\omega^2+r^{4\beta}k_y^2+m^2)R(r)=\lambda R(r)r^{2\beta},
\ee
with separation constant $\lambda$. These equations admit an infinite set of solutions labeled by the eigenvalue $\lambda$. The solutions  must satisfy the boundary conditions $R(\infty)=X(\pm\infty)=0$ (regularity at the horizon and square-integrability along $x$), and $R(r\to \epsilon)=1$, with $\epsilon$ a UV cutoff. 

\subsubsection{The $X(x)$ equation of motion}
First consider the equation for $X(x)$--this has solutions which are independent of both the values of the scaling exponents in the metric and the scalar mass. Defining $\tilde x=(k_yx+k_z)/\sqrt{k_y}$ and $\lambda_n=\lambda/k_y$ we get the equation,
\be
X_n''(\tilde x)-(\tilde x^2-\lambda_n)X_n(\tilde x)=0.
\ee
This is the equation of a quantum harmonic oscillator with $ (m\omega)^2=1$ and energy eigenvalues $E_n=\lambda_n/2m=\omega(n+1/2)$. This equation has the familiar solutions
\be\label{eq:Xsol}
X_n(\tilde x)=e^{-\frac{\tilde x^2}{2}}H_{(\lambda_n-1)/2}(\tilde x)
\ee
for $\lambda_n=2n+1,~~n=0,1,2,\ldots\infty,$  where $H_\nu(x)$ are the Hermite polynomials and $c_n$ are constants. 

The $X(x)$ part of Green's function will be a specific linear combination of the set, $\{X_n\}$, of solutions we just found, such that it satisfies the boundary condition:  $X(x)=\delta(x-x')$ when $r\to 0$. Because the $r$ and $x$ equations decouple, we simply have that $X(x)$ itself is a $\delta$-function in $(x-x')$, independent of $r$. We will build up this $\delta$-function using the basis we found in equation (\ref{eq:Xsol}),
\be\label{eq:deltafunction}
X(x)=\sum_{n=0}^\infty c_nX_n(\tilde x)=\delta(x-x'),
\ee  
where the $c_n$ are constants. We will solve for the $c_n$ by using the orthogonality relations for Hermite polynomials,
\be
\int_{-\infty}^\infty dx e^{-x^2}H_m(x)H_n(x)=2^m m!\sqrt{\pi}\delta_{mn}.
\ee
Therefore, in order to satisfy equation (\ref{eq:deltafunction}), we require that\footnote{In solving for the $c_n$, we have used the identity $\delta(\alpha x)={\delta(x)\over \alpha}$, which allows us to write, $\delta(x-x')=\sqrt{k_y}\delta(\tilde x-\tilde x')$.}
\be\label{eq:cn}
c_n={1\over\sqrt{k_y\pi}}\frac{1}{2^nn!}e^{-{\tilde x'^2\over 2}}H_n(\tilde x')
\ee
with $\tilde x'=(k_yx'+k_z)/\sqrt{k_y}$. This solution fully captures the $x$ and $k_z$-dependence of the Green's function, independent of our choice of $\beta,\beta_t$, and $m$. There will be additional dependence on $\omega$ and $k_y$ which will enter through the equation for $R(r)$. We turn to this in the next section.

\subsubsection{The radial equation, $R(r)$}
Now let's consider the radial equation. We can write equation (\ref{eq:req}) as
\be
r^2R_n''(r)-4\beta rR_n'(r)-(r^{2\beta_t}\omega^2+r^{2\beta}k_y\lambda_n+r^{4\beta}k_y^2+m^2)R_n(r)=0,
\ee
which is not solvable for general $\beta, \beta_t$, but which we will solve for a few specific parameter values. First we set $\beta_t=1$. The following are solutions for selected values of $\beta$,
\bea\nonumber
R_n(r)&=&\frac{\sqrt{r\omega}}{\sqrt{\epsilon\omega}}\frac{K_{\frac{1}{2}\sqrt{1+4k_y\lambda_n+4k_y^2+4m^2}}(r\omega)}{K_{\frac{1}{2}\sqrt{1+4k_y\lambda_n+4k_y^2+4m^2}}(\epsilon\omega)},~~~\beta=0\\\nonumber
R_n(r)&=&4e^{-kr}(kr)^3\Gamma(A_n ) U(A_n,4,2kr ),~~~\beta=1/2,~m=0\\\label{eq:Rsols}
R_n(r)&=&\frac{e^{-kr}(2kr)^{1+{B\over 2}}U\left(A_n-2+{B\over 2},B,2kr\right)}{e^{-k\epsilon}(2k\epsilon)^{1+{B\over 2}}U\left(A_n-2+{B\over 2},B,2k\epsilon\right)},~~~\beta=1/2,~m\neq 0,
\eea
where the parameters are defined as,
\be\nonumber
k=\sqrt{\omega^2+k_y^2},~A_n=2+\frac{k_y\lambda_n}{2k},~
B=1+\sqrt{9+4m^2},
\ee
and $K_\nu(x)$ is the modified Bessel function of the second kind and $U(a,b,x)$ is the confluent hypergeometric function of the second kind.
We have already imposed the regularity condition at the horizon, $R(\infty)\to 0$ and the normalization condition $R(r\to\epsilon)\to 1.$ (It turns out that for the case of $\beta=1/2, m=0$ we do not need a UV regulator $\epsilon$ to satisfy the normalization condition.)

Therefore, the full Green's function is
\be
g(r,\vec k,x,x')=\sum_{n=0}^\infty c_ne^{-\frac{\tilde x^2}{2}}H_{n}(\tilde x)R_n(r)
\ee
with $c_n$ as given in equation (\ref{eq:cn}). Since $R_n(r)\to 1$ at the boundary, this reduces to the delta function in $x,x'$ of equation (\ref{eq:deltafunction}) at $r\to\epsilon$ as required. When we plug this solution back into the action we get the momentum space two point function of equation (\ref{eq:full2ptfn}), which becomes
\bea\nonumber
\langle\mathcal O(x,\vec k)\mathcal O(x',-\vec k)\rangle&=& \sqrt{-g}g^{rr}\sum_{n=0}^\infty c_n e^{-\frac{\tilde x^2+\tilde x'^2}{2}}H_{n}(\tilde x)\partial_r R_n(r,\vec k)|_{r\to \epsilon}\\\label{eq:Nexp}
&=& \sum_{n=0}^\infty\frac{1}{\sqrt{k_y\pi}2^nn!}e^{-\frac{\tilde x^2+\tilde x'^2}{2}}H_{n}(\tilde x)H_n(\tilde x')\frac{\partial_r R_n(r,\vec k)}{r^{4\beta}}\Big|_{r\to \epsilon}
\eea
Note that the $x$-dependence of equation (\ref{eq:Nexp}) is a weighted sum of harmonic oscillator eigenfunctions; this is a coherent state of a quantum harmonic oscillator. In order to get the $\omega$ and $k_y$-dependence, we will need to consider the expansion of $R(r)$ near $r=\epsilon$. 

\subsection{Field theory two-point function}
Now we will evaluate equation (\ref{eq:Nexp}) to get the position space two-point function, $\langle\mathcal O(\tau, \vec x)\mathcal O(\tau', \vec x')\rangle$. We will do this for each of the three solutions for $R_n(r)$ found above. We also discuss the interpretation in both the KK-reduced bulk and the dual field theory.
\subsubsection{$\beta=0$: no spatial scaling}
First we examine the case in which $\beta=0$; the three spatial directions decouple from the $r, t$ part of the metric. In this case we will need to expand the modified Bessel function around $r=\epsilon$,
\be
\frac{\partial_r R_n(r,k)}{r^{4\beta}}\Big|_{r\to \epsilon}\sim\lim_{r\to\epsilon}\frac{\partial}{\partial_r}\left (\frac{\sqrt{r}}{\sqrt{\epsilon}}\frac{K_\nu(r\omega)}{K_\nu(\epsilon\omega)}\right )
\ee
where $\nu =\frac{1}{2}\sqrt{4k_y^2+1+4k_y(2n+1)+4m^2}$.

We will make use of the expansion
\be
K_\nu(\alpha r)\sim 2^{\nu-1}\Gamma(\nu)(\alpha r)^{-\nu}(1+\ldots)-2^{-\nu-1}\frac{\Gamma (1-\nu)}{\nu}(\alpha r)^\nu (1+\ldots),
\ee
where the $\ldots$  are powers of $(\alpha r)^{2n}, n=1,2,\ldots$, where the second term in the expression is the leading order non-analytic piece in the action. We will also make use of the limit
\be\label{eq:BesselKlimit}
\lim_{r\to\epsilon}\frac{\partial}{\partial_r}\left (\frac{(\omega r)^{1/2}}{(\omega \epsilon)^{1/2}}\frac{K_\nu(\omega r)}{K_\nu(\omega \epsilon)}\right )\sim \frac{2}{\epsilon}\frac{\Gamma (1-\nu)}{\Gamma (\nu)}\left (\frac{\omega\epsilon}{2}\right )^{2\nu},
\ee
which gives us the leading order behavior of the two-point function in momentum space. The above expression makes it clear that the $n=0$ term will dominate in equation (\ref{eq:Nexp}). This is just the Hermite polynomial $H_0$, which is constant. Keeping only this term, we get that the full two-point function in momentum space is
\be
\langle\mathcal O(x,\vec k)\mathcal O(x',-\vec k)\rangle\sim\int dxdx'{2\over \epsilon}{1\over\sqrt{k_y\pi}}\frac{\Gamma(1-\nu)}{\Gamma(\nu )}e^{-\frac{\tilde x^2+\tilde x'^2}{2}}\left (\frac{\omega\epsilon}{2}\right )^{2\nu}.
\ee
Expanding the exponent in the above equation, we see that the $k_z$ dependence of the momentum space propagator is $\langle\mathcal O(x,\vec k)\mathcal O(x',-\vec k)\rangle\sim e^{-k_z(x+x')-{k_z^2\over k_y}}$. When we Fourier transform this piece back into position space, we will make use of the integral
\be\label{eq:zint}
\int dk_z e^{ik_z(z-z')-k_y(x^2+x'^2)-k_z(x+x')-{k_z^2\over k_y}}=\sqrt{\pi k_y}e^{-{k_y\over 4}(\Delta x^2+\Delta z^2-2i(x+x')|\Delta z|)},
\ee
where $\Delta x=x-x'$.
Now we will perform the integral over $k_z$ using equation (\ref{eq:zint}) and the integral over $\omega$
to get
\be\label{eq:prop1}
\langle\mathcal O(\tau,x,z,k_y)\mathcal O(\tau',x',z',-k_y)\rangle\sim-\left({2\over \epsilon}\right)^{1-2\nu}\frac{\sqrt{2\pi}}{|\Delta\tau|^{2\nu+1}}\frac{\Gamma(2\nu+1)}{\Gamma(\nu)^2}e^{-{k_y\over 4}(\Delta x^2+\Delta z^2-2i(x+x')|\Delta z|)},
\ee
where, as a reminder, $\nu={\sqrt{(2k_y+1)^2+4m^2}\over 2}$. Since in this case we chose $\beta=0$, the spatial directions are not scale-invariant, and the $t$ and $r$ directions form an $AdS_2$. The dependence of the two-point function on time scales as a power law with dimension $\Delta_\tau=\nu+{1\over 2}=\frac{1}{2}(1+\sqrt{4k_y^2+1+4k_y+4m^2})$, which depends on the values of $k_y$ and $m$. This is reminiscent of the fact that the two-point function of a scalar in $AdS_2\times R^2$ has a momentum-dependent dimension\cite{ads2}. 

For a particular fixed value of (nonzero) $k_y$, the correlation function decays exponentially in $\Delta x^2$ and $\Delta z^2$.\footnote{Of course, if $k_y=0$ the dependence on $y$ is removed and the result is the two-point function for a massive scalar in an $AdS_2\times R^2$ geometry \cite{ads2}.} We find that the $x$ and $z$ dependence is independent of mass. Note that, up to a phase, we see explicitly that $\langle\mathcal O\mathcal O\rangle$ only depends on $|\Delta x|$, and translation invariance in the $x$ direction is restored! In fact, the $x$ and $z$ dependence is similar to that of a propagator for a harmonic oscillator in its ground state. Note that the form of the Green's function for $\phi$ is non-universal; for example, the the $x$ and $z$-dependence of equation (\ref{eq:gauge_green}) reflects our choice of metric gauge. Under a gauge transformation, the form of the Landau-level basis transforms, and this would also cause the form of the bulk propagator to transform. In addition, the two-point function of a scalar operator should depend upon the state in the field theory in which it is evaluated. However, given a fixed state in the dual field theory, the result for the two-point function of $\mathcal O$ should be independent of our gauge choice in the bulk. This is exactly what we find in equation (\ref{eq:prop1}) for a scalar in the probe limit. We would also get corrections to equation (\ref{eq:prop1}) by considering the higher-order Hermite polynomials, however these terms are subleading for $n\geq 1$. However, we will find that these terms are no longer subleading when we consider the two-point function for $\beta={1\over 2}$ below.

\subsubsection{Four-dimensional interpretation}
 In the solution we found, $k_y$ can take on a continuum of values. However, if we were to KK-reduce along the $y$-direction like in \S~\ref{sec:KK}, then $k_y$ would take on a discretuum of values, $k_y\sim 2\pi nR$ with $n\in\mathbb{Z}$ and $R$ the radius of the compactification circle. What happens to $\phi$, a neutral scalar in five dimensions, after compactification? Since $k_y$ is now no longer continuous, this results in a discrete tower of scalar operators in four dimensions; these operators are charged under the four-dimensional magnetic field coming from the metric mode $B_\mu dx^\mu$, and have charged labeled by their value of $k_y$, the momentum along the compactified direction. Thus in five dimensions, $\phi$ is neutral but couples to the gauge field via the metric, whereas upon reduction to four dimensions, $\phi$ produces one neutral scalar (the zero mode, $k_y=0$), and an infinite number of charged scalar operators coupled to the four-dimensional magnetic field. We see from equation (\ref{eq:prop1}) that the two point function for scalars with larger fixed $k_y$ values, and thus larger charge, decays faster exponentially, as $\sim e^{-k_y(\Delta x^2+\Delta z^2)}$. 

It is quite interesting that when we choose $\beta_x=\beta_z=0$ in the KK reduction solution of \S~\ref{sec:KK}, we get a four-dimensional metric that is $AdS_2\times R^2$, with no hyperscaling violation exponent, and an action with a massless gauge field which produces a $B$-field in the $r$-direction. This is exactly the near-horizon limit of a magnetically-charged, extremal Reissner-Nordstrom (RN) black brane. Though we computed (\ref{eq:prop1}) using the five-dimensional type-II metric, for a fixed $k_y$ value, two-point function in equation (\ref{eq:prop1}) is that of a four-dimensional scalar charged under this magnetic field. The nice thing about the five-dimensional computation is that, already in the probe limit, because of the anisotropic metric, we can  explicitly see the appearance of Landau levels in the solution of a charged particle in a constant magnetic field. 

We have seen the appearance of Landau levels in magnetic $AdS_2\times R^2$ geometries before \cite{vlattice}. In this case, a charged scalar field is unstable to the production of a vortex lattice. This lattice is composed of an infinite number of Landau level solutions, periodically overlaid. Therefore, we may expect type-II geometries may be unstable to vortex lattice phases in the IR as well. It would be interesting to investigate two-point functions for more interesting states such as those in \cite{vlattice} after including the backreaction of the bulk solution on the metric. It is also interesting that these magnetic $AdS_2\times R^2$ solutions are S-dual to solutions with an electric field, which have non-perturbative instabilities due to monopole operators\cite{sachdev}.

\subsubsection{$\beta=1/2$, $m=0$: massless scalars with scaling}
In this case, the expansion of $R_n(r)$ around $r=0$ is,
\bea\nonumber
R_n(r)&=&1+\ldots+\frac{(kr)^3}{3}\Big (8-{161A_n\over 6}+19A_n^2-{11A_n^3\over 3}\\\nonumber
&&+4\gamma+2\log(2kr)+2\psi_0(A_n)\Big)+\ldots
\eea
where $\gamma$ is the Euler-Mascheroni constant,  $\psi_0(x)$ is the digamma function, and, as before, $k=\sqrt{k_y^2+\omega^2}$ and $A_n=2+\frac{k_y\lambda_n}{2k}$. 
When we plug this back into equation (\ref{eq:Nexp}) to get the two-point function, we will need to pick out the leading non-analytic terms in $k_y, \omega$ in the expansion
\bea\nonumber
\lim_{r\to\epsilon}{1\over r^2}\frac{\partial}{\partial r}R_n(r)&\sim&\frac{k^3}{3}(4\gamma+1+2\log(2k)+2\psi_0(A_n))\\\label{eq:Rn}
&&+\frac{31}{3\cdot 12}k^2k_y\lambda_n-{1\over 4}kk_y^2\lambda_n^2-\frac{11}{3\cdot 24}k_y^3\lambda_n^3
\eea
There is additional $k_y$ dependence coming from the $x$-dependent part of the propagator.

Because the scaling with radius is independent of Landau level, $n$, all the Hermite polynomials will contribute at the same order in the limit as $r\to 0$. The Fourier transform in $k_z$ of the spatial part in (\ref{eq:Nexp}) has the form 
\be\label{eq:xz}
\int dk_z e^{ik_z(z-z')}\frac{1}{\sqrt{k_y}}e^{-\frac{\tilde x^2+\tilde x'^2}{2}}H_{n}(\tilde x)H_n(\tilde x')\sim e^{-{k_y\over 4}(\Delta x^2+\Delta z^2-2i(x+x')|\Delta z|)}\prod_{i=1}^n (k_y(\Delta x^2+\Delta z^2) +a_i),
\ee
where $a_i$ is an integer and the $n$th term is suppressed by ${1\over 2^n n!}$ as in (\ref{eq:Nexp}). It is satisfying that the spatial dependence remains independent of the metric gauge we have chosen and isa function of $|\Delta x|$ and $|\Delta z|$ only. Fourier transforming equation (\ref{eq:Rn}) in $\omega$ results in contact terms and terms with inverse power law scaling $\sim |\Delta \tau|^{-2},|\Delta \tau|^{-1}$. As we have discussed in the previous section, we should think of $k_y$ as fixed, as it becomes the charge of the scalar and only takes discrete values in four dimensions.

\subsubsection{$\beta=1/2$, $m\neq 0$: massive scalars with scaling}
Finally, we'll see how much we can deduce about the two-point function of massive scalars in a type-II background with scaling. First we expand the numerator of the third solution in equation (\ref{eq:Rsols}) around $r=0$,
\be
R_n(r)\sim(2kr)^{2-{B\over 2}}\frac{\Gamma(B-1)}{\Gamma\left (A_n+{B\over 2}-2\right )}(1+\cdots)+(2kr)^{1+{B\over 2}}\frac{\Gamma(1-B)}{\Gamma\left (A_n-1-{B\over 2}\right )}(1+\cdots).
\ee
The terms written are the leading order non-analytic terms in $\omega$ and $k_y$; the $\ldots$ give higher order terms, while the dependence on $k_z$ enters only through the exponential and Hermite polynomial terms. Plugging back in and taking the limit as $r\to\epsilon$, we get the full form of the propagator in momentum space,
\bea\nonumber
\langle\mathcal O(x,\vec k)\mathcal O(x',-\vec k)\rangle&=&-\epsilon^{\sqrt{9+4m^2}-3} \left ({3+\sqrt{9+4m^2}\over 2}\right)\sum_{n=0}^\infty\frac{1}{\sqrt{k_y\pi}2^nn!}e^{-\frac{\tilde x^2+\tilde x'^2}{2}}H_{n}(\tilde x)H_n(\tilde x')\\&&\times \frac{\Gamma(-\sqrt{9+4m^2})\Gamma\left ({k_y\lambda_n\over 2k}+{1\over 2}+{\sqrt{9+4m^2}\over 2}\right )}{\Gamma(\sqrt{9+4m^2})\Gamma\left ({k_y\lambda_n\over 2k}+{1\over 2}-{\sqrt{9+4m^2}\over 2}\right )}\left(2\sqrt{\omega^2+k_y^2}\right)^{\sqrt{9+4m^2}},
\eea
which is quite intricate, with particularly complicated dependence on the momentum $k_y$. However, this has the scaling behavior of an operator of dimension $\Delta=3+\sqrt{9+4m^2}$.\footnote{This is true except in the special case $9+4m^2= n^2$, where $n$ is an integer.} The $x$ and $z$ dependence will be as in equation (\ref{eq:xz}).

\section{Discussion}\label{sec:discussion}
We have explored many aspects of homogeneous, anisotropic three-dimensional brane horizons, including the behavior of two-point functions for general type I geometries with anisotropic spatial scaling, and the occurrence of tidal force singularities in general Bianchi-type spacetimes.  Our specific emphasis has been on the type II case, which has symmetries of the Heisenberg algebra. We have generalized such spacetimes to include hyperscaling violation in five dimensions, and explored the relationship to four-dimensional bulk geometries with a magnetic field via KK-reduction. In addition, we've computed scalar two-point functions in the probe limit and found that they have the structure of a theory with Landau levels.

There are a number of interesting directions one could explore with respect to the ``phenomenology" of spacetimes with Heisenberg symmetry. One natural extension would be to consider generalizations of type II metrics (or other Bianchi types) to nonzero temperature, which would involve adding an emblackening factor to the metric. Then one could investigate the temperature dependence of thermodynamic quantities and transport of systems with a magnetic field, purely through analysis of the metric. Another generalization would be to add a five dimensional compact manifold which scales with radius, which may allow one to have more general values of the critical exponents in the noncompact geometry, or to allow the metric gauge field, $B_\mu$ to scale with radius. One could also consider including the radial coordinate in the algebra, like in some of the geometries discussed in \cite{bigteam2}.

Another consideration would be to study the IR of such theories, including potential instabilities to the formation of inhomogeneous phases which appear in theories with Landau levels. One may be able to find solid, insulating phases which are analogues of states that appear in the two-dimensional electron gas. By exploiting the connection between the five-dimensional and KK-reduced theories, one could compute transport properties in (2+1)-dimensional magnetic systems which come from the five-dimensional metric gauge field. It would also be very interesting to explore the relationship of type II geometries to quantum Hall physics, which has been studied in holography in \cite{fqhe}.

Finally, one would like to construct a UV completion of type II theories which asymptote to AdS space. So far we have only been working at the level of an effective theory, but it's natural to expect such theories could be constructed via string compactifications with fluxes. The constructions we have discussed which support type II involve a massive gauge field, which can be produced from KK reduction of a higher dimensional theory. However, this makes finding a full ten dimensional solution particularly difficult, as one then has to deal with a whole tower of KK modes which arise from the compact dimensions. If there was another way to support type II, perhaps with axions or running scalar fields, then one may be able to construct these from a ten dimensional theory with fluxes on a small compact Calabi-Yau. This method is being applied to produce bulk geometries with $AdS_2$ and hyperscaling violation metrics \cite{huajia}. Having a full ten dimensional solution from string theory would be a natural way to investigate the RG flow and stability of systems with type II isometries in detail, and the connection between five-dimensional type-II geometries and (2+1)-dimensional field theories with a constant magnetic field.

\bigskip
\centerline{\bf{Acknowledgements}}

The author would particularly like to thank S. Kachru and H. Wang for many useful conversations. Additional thanks to X. Dong and G. Torroba for discussions, and N. Bao, N. Paquette, and G. Torroba for comments on a draft. The author is supported in part by the John Templeton Foundation.

\appendix

\section{Metrics, curvature, null energy conditions}\label{sec:NEC}

Here we write out the Bianchi type metrics for reference, including their scalar curvature and null-energy conditions in full which are referred to in \S~\ref{sec:type1} and \S~\ref{sec:scaling}.
\subsection{Type I}
Section \ref{sec:type1} discuss properties of general type I spacetimes. The most general Type I metric including a hyperscaling violation exponent $\theta$ is
\be
ds^2=L^2r^{2\theta/3}(-r^{-2\beta_t}dt^2+{dr^2\over r^2} +r^{-2\beta_x}dx^2+dr^{-2\beta_y}dy^2 +r^{-2}dz^2),
\ee
noting that we can always rescale $r$ such that one of the $x_i$s scales with weight 1.
The Ricci scalar is
\be
R=-r^{-2\theta/3}\left (\frac{2(\beta_t^2+\beta_x^2+\beta_y^2+\beta_t(\beta_x+\beta_y)+\beta_x\beta_y+\beta_x+\beta_y+\beta_t)}{L^2}-\frac{8\theta (1+\beta_t+\beta_x+\beta_y)}{3L^2}+\frac{4\theta^2}{3L^2}\right )
\ee
which reduces to
\be
R=-\frac{2(\beta_t^2+\beta_x^2+\beta_y^2+\beta_t(\beta_x+\beta_y)+\beta_x\beta_y+\beta_x+\beta_y+\beta_t)}{L^2}
\ee
for $\theta=0$.

The null vector is $N^\mu=(r^{\beta_t},c_r r,c_1r^{\beta_x},c_2r^{\beta_y},c_3r).$ We get four inequalities from choosing the four vectors such that $c_i=1$, $c_{j\neq i}=0$ and computing $T_{\mu\nu}N^{\mu}N^{\nu}\geq 0$ for each vector. The conditions on the $\beta$s are then
\bea\nonumber
\beta_t(\beta_x+\beta_y+1)-\beta_x^2-\beta_y^2-1 -\beta_t\theta+{\theta^2\over 3}&\geq &0\\\nonumber
(\beta_t-1)(\beta_t+\beta_x+\beta_y+1-\theta)&\geq &0\\\nonumber
(\beta_t-\beta_x)(\beta_t+\beta_x+\beta_y+1-\theta)&\geq &0\\
(\beta_t-\beta_y)(\beta_t+\beta_x+\beta_y+1-\theta)&\geq &0
\eea
which reduce to
\bea\nonumber
\beta_t(\beta_x+\beta_y+1)-\beta_x^2-\beta_y^2-1 &\geq &0\\\nonumber
(\beta_t-1)(\beta_t+\beta_x+\beta_y+1)&\geq &0\\\nonumber
(\beta_t-\beta_x)(\beta_t+\beta_x+\beta_y+1)&\geq &0\\
(\beta_t-\beta_y)(\beta_t+\beta_x+\beta_y+1)&\geq &0
\eea
for $\theta=0$.
\subsection{Type II}
Here we will give the null energy conditions for the metrics explored in \S~\ref{sec:scaling}. The most general type II metric which includes the possibility of a hyperscaling violation exponent is
\be
ds^2=L^2r^{2\theta/ 3}\left (-r^{-2\beta_t}dt^2+{dr^2\over r^2}+r^{-2\beta_x}dx^2+r^{-2\beta_z}dz^2+r^{-2(\beta_x+\beta_z)}(dy-xdz)^2\right ).
\ee
The Ricci scalar for this metric is 
\be
R=r^{-2\theta/3}\left (-\frac{2(\beta_t^2+2\beta_t(\beta_x+\beta_z)+3(\beta_x^2+\beta_z^2)+5\beta_x\beta_z)}{L^2}+\frac{8\theta(\beta_t+2(\beta_x+\beta_z))-4\theta^2}{3L^2}-\frac{1}{2L^2}\right ),
\ee
which is not constant and diverges as $r\to 0$. However, if we set $\theta=0$ it becomes
\be
R=-\frac{2(\beta_t^2+2\beta_t(\beta_x+\beta_z)+3(\beta_x^2+\beta_z^2)+5\beta_x\beta_z)}{L^2}-\frac{1}{2L^2}
\ee
which is constant. In fact, all curvature invariants are finite for all Bianchi metrics without hyperscaling violation.

The null vectors are given by $N^\mu=(r^{\beta_t},c_r r,c_1r^{\beta_x},c_2r^{\beta_x+\beta_z}+c_3xr^{\beta_z},c_3r^{\beta_z})$ where the $c_i$s are constants. For nonzero $\theta$, the constraints are
\bea\nonumber
\beta_t(\beta_x+\beta_z)-(\beta_x^2+\beta_z^2+\beta_x\beta_z)-{\beta_t\theta\over 2}+{\theta^2\over 6}&\geq &0\\\nonumber
\beta_t^2+\beta_t\beta_x+2\beta_t\beta_z-2\beta_x\beta_z-2\beta_x^2+\theta(\beta_x-\beta_t)-{1\over 2} &\geq & 0\\\nonumber
\beta_t^2+2\beta_t\beta_x+\beta_t\beta_z-2\beta_x\beta_z-2\beta_z^2+\theta(\beta_z-\beta_t)-{1\over 2}&\geq &0\\
\beta_t^2+\beta_t\beta_x-2\beta_x^2+\beta_t\beta_z-4\beta_x\beta_z-2\beta_z^2+\theta(\beta_x+\beta_z-\beta_t)+{1\over 2} &\geq & 0
,
\eea
which reduce to
\bea\nonumber
\beta_t(\beta_x+\beta_z)-\beta_x^2-\beta_z^2-\beta_x\beta_z&\geq &0\\\nonumber
\beta_t^2+\beta_t\beta_x+2\beta_t\beta_z-2\beta_x^2-2\beta_x\beta_z-{1\over 2} &\geq & 0\\\nonumber
\beta_t^2+\beta_t\beta_z+2\beta_t\beta_x-2\beta_z^2-2\beta_x\beta_z-{1\over 2} &\geq & 0\\
\beta_t^2+\beta_t (\beta_x+\beta_z)-2(\beta_x+\beta_z)^2+{1\over 2}&\geq &0
\eea
for $\theta=0$.

%\newpage
%%%%%%%%%%%%%%%%%%%%%%%%%%%%%%%%%%%%%%%%%%%%%%%%%%%%%%%%%%%%
%%%%%%%%%%%%%%%%%%%%%%%%%%%%%%%%%%%%%%%%%%%%%%%%%%%%%%%%%%%%

\bibliographystyle{JHEP}
\renewcommand{\refname}{Bibliography}
\addcontentsline{toc}{section}{Bibliography}
\providecommand{\href}[2]{#2}\begingroup\raggedright

\end{document}